\documentclass[
  aps,prl,
  reprint,
  superscriptaddress,
  amsmath,amssymb,
  floatfix,
  nofootinbib
]{revtex4-2}

\usepackage{graphicx}
\usepackage{bm}
\usepackage{xcolor}
\usepackage{hyperref}
\hypersetup{colorlinks=true,
            linkcolor=blue!55!black,
            citecolor=blue!55!black,
            urlcolor=blue!55!black}
\renewcommand{\selectlanguage}[1]{}
\usepackage{amsthm}
\usepackage[capitalize,nameinlink]{cleveref}

\theoremstyle{plain}
\newtheorem{theorem}{Theorem}
\newcommand{\PP}{\mathbb P}
\newcommand{\EE}{\mathbb E}
\newcommand{\ZZ}{\mathbb Z}
\newcommand{\one}{\mathbf 1}

\begin{document}

\title{Onset of Ferromagnetic Order in Ising Chains}

\author{Qihang Wang}
\affiliation{School of Mathematical Sciences, Peking University, Beijing 100871, China}
\author{Zhiyi Li}
\affiliation{Department of Modern Physics, University of Science and Technology of China, Hefei, Anhui 230026, China}
\affiliation{CAS Key Laboratory of Theoretical Physics, Institute of Theoretical Physics, Chinese Academy of Sciences, Beijing 100190, China}
\author{Kun Chen}
\email[Corresponding author: ]{chenkun@itp.ac.cn}
\affiliation{Institute of Theoretical Physics, Chinese Academy of Sciences, Beijing 100190, China}
\author{Zhiyuan Yao}
\email[Corresponding author: ]{yaozy@lzu.edu.cn}
\affiliation{Lanzhou Center for Theoretical Physics, Key Laboratory of Theoretical Physics of Gansu Province, Key Laboratory of Quantum Theory and Applications of MoE, Gansu Provincial Research Center for Basic Disciplines of Quantum Physics, Lanzhou University, Lanzhou 730000, China}

\date{\today}

\begin{abstract}
The onset of ferromagnetic order in one dimension reflects a competition between domain energy and entropy. For Ising chains with arbitrary translation-invariant, nonnegative, summable pair interactions, we establish a necessary and sufficient condition for an ordered phase at finite temperature: the energy required to reverse an interval of $N$ spins must be asymptotically bounded below by a positive multiple of $\ln N$. When this energy is also asymptotically bounded above by a multiple of $\ln N$, we prove that throughout the ordered phase the spontaneous magnetization is bounded below by a positive value proportional to the square root of the temperature, so that the magnetization jumps to a nonzero value at the critical temperature. This extends the proof of the magnetization jump beyond pointwise inverse-square bounds, to sparse and irregular interactions for which $r^2J(r)$ is unbounded.
\end{abstract}

\maketitle

\textit{Introduction}.---Order results from a competition between energy and entropy, and in a one-dimensional Ising chain with finite-range interactions entropy always wins~\cite{Ising_1925bz,Peierls_1936oi,Landau_1980sp}. Long-range couplings~\cite{Fisher_1972ce,Defenu_2023lr} change this balance through the domain energy, which measures the cost of reversing an interval of $N$ spins in an ordered background. This energy grows without bound only if the bonds at large distances are strong enough that the first moment $\sum_r rJ(r)$ diverges; couplings with a finite first moment keep it bounded and never order~\cite{Ruelle_1968sm}. For power laws $J(r)\propto r^{-\alpha}$, order was conjectured for $1<\alpha<2$~\cite{Kac_1969cb} and proved by Dyson~\cite{Dyson_1969eo}. At the marginal inverse-square law, energy and entropy both grow as $\ln N$, and the outcome depends on temperature; Fr\"ohlich and Spencer proved that this chain orders at low temperature~\cite{Frohlich_1982tp,Imbrie_1982do}. Beyond power laws, order has been proved under regular decay, pointwise inverse-square lower bounds, reflection positivity, or sufficiently strong weighted isoperimetry~\cite{Cassandro_2005go,Bissacot_2018cm,Affonso_2024pt,Gracar_2025fo,Newman_1986od,Duminil-Copin_2024lr,Frohlich_1978pt,Hartarsky_2026wi}. Conversely, nonordering criteria, culminating in a uniqueness theorem of Berbee, exclude order whenever the domain energy grows too slowly compared with $\ln N$~\cite{Dyson_1969ne,Rogers_1981ao,Simon_1981re,Berbee_1989uo}, and the same energy controls uniqueness on a half line~\cite{Fernandez_2007ua,JOHANSSON_2017pt}. A general coupling, sparse or irregular, need not decay as a power law, and whether the domain energy alone decides if it orders has remained open.

Even where order exists, the magnetization need not appear continuously. For the inverse-square chain, Thouless predicted that the magnetization jumps to a nonzero value at the transition~\cite{Thouless_1969lr}. Continuity of the transition has been proved for long-range percolation and Ising chains whose couplings decay more slowly than $r^{-2}$~\cite{Berger_2002tr,Aizenman_2015rc,Hutchcroft_2021pl} and in other geometries~\cite{Bricmont_1986ot,Raoufi_2016an,Hutchcroft_2023co}, but none of these arguments reaches the inverse-square law. Discontinuous order appeared in a hierarchical model~\cite{Dyson_1971ai}, and the transition was likened to Kosterlitz--Thouless defect unbinding in a gas of domain walls~\cite{Anderson_1969er,Anderson_1970er,Anderson_1971sn,Kosterlitz_1973om,Kosterlitz_1976pt,Cardy_1981od,Bhattacharjee_1981sp}. Through a continuum version of the inverse-square chain, the same jump governs the localization transition of the Ohmic spin-boson model~\cite{Chakravarty_1982qf,Leggett_1987do}. The jump was proved first assuming a finite truncated susceptibility~\cite{Simon_1981re}, then unconditionally by percolation, for long-range percolation with inverse-square bond probabilities~\cite{Aizenman_1986do} and for Ising chains with $r^2J(r)$ bounded~\cite{Aizenman_1988do}; an intermediate ordered phase~\cite{Imbrie_1988ai} and a new multiscale proof of the jump for the pure inverse-square coupling $J(r)\propto r^{-2}$~\cite{Duminil-Copin_2024lr} followed. All these proofs bound $J(r)$ at every distance, and whether a logarithmic bound on the domain energy alone forces the magnetization to jump has remained open.

In this Letter, we answer both questions for arbitrary translation-invariant, nonnegative, summable Ising couplings. Unlike pointwise hypotheses, which constrain $J(r)$ at every distance, the domain energy sums the couplings over all distances and is equally defined for sparse and irregular couplings. First, order exists at some finite temperature if and only if the domain energy is asymptotically bounded below by a positive multiple of $\ln N$. Necessity follows from the uniqueness theorem of Berbee~\cite{Berbee_1989uo}; we prove sufficiency. Second, when the domain energy is also asymptotically bounded above by a multiple of $\ln N$, the spontaneous magnetization throughout the ordered phase is bounded below by a positive value set by the temperature, so it jumps to a nonzero value at the critical point. This class contains the inverse-square couplings, for which we recover the known bound, and sparse chains with bonds only at the distances $2^k$, for which $r^2J(r)$ is unbounded.

\begin{figure}[t]
\centering
\includegraphics[width=\columnwidth]{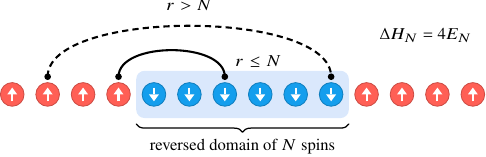}
\caption{Energy of a reversed domain. Red (blue) disks are up (down) spins. A reversed domain of $N$ spins in a polarized background breaks both short and long bonds across its walls. There are exactly $2\min(r,N)$ broken bonds of length $r$, so $\Delta H_N=4E_N$.}
\label{fig:domain}
\end{figure}

\textit{Model and domain energy}.---We consider Ising spins $\sigma_i=\pm1$ on a chain with a ferromagnetic, translation-invariant pair coupling,
\begin{equation} \label{eq:model}
	H=-\sum_{i<j}J(j-i)\,\sigma_i\sigma_j,\qquad J(r) \ge 0,
\end{equation}
that is summable, $\Sigma_J=\sum_{r\ge1}J(r)<\infty$. Nothing else is assumed: $J$ need not be monotone or decay as a power law, may vanish at most distances, and need not include a nearest-neighbor bond. We shall set the lattice spacing and $k_{\rm B}$ to be one. An infinite-volume Gibbs measure at inverse temperature $\beta$ is a probability measure on spin configurations of the whole chain whose conditional distribution in every finite interval, given the spins outside, is the Boltzmann distribution $\propto e^{-\beta H}$; each such measure describes an equilibrium state of the infinite chain, and an ordered phase has at least two, obtained with all spins up and all spins down outside. The spontaneous magnetization $m_+(\beta)=\langle\sigma_0\rangle_\beta^+$ is the magnetization of the Gibbs measure obtained as the infinite-volume limit with all spins up outside, and $\beta_c=\inf\{\beta:m_+(\beta)>0\}$. To quantify the energy cost of a reversed domain, consider flipping $N$ consecutive spins in the fully polarized state. As illustrated in Fig.~\ref{fig:domain}, this breaks exactly $2\min(r,N)$ bonds of length $r$, each at an energy cost $2J(r)$. The energy of the reversed domain is therefore
\begin{equation}
\Delta H_N=4E_N,\qquad E_N=\sum_{r\ge1}\min(r,N)\,J(r).
\end{equation}
Here $E_N$ is the total coupling across one of the two walls, and we call it the domain energy; bonds shorter than the domain enter with their length, longer ones with the length $N$ of the domain. For irregular interactions, $E_N/\ln N$ need not converge; its asymptotic lower and upper envelopes are
\begin{equation}
C_-=\liminf_{N\to\infty}\frac{E_N}{\ln N},\qquad C_+=\limsup_{N\to\infty}\frac{E_N}{\ln N},
\end{equation}
with $0\le C_-\le C_+\le\infty$, both in units of energy. For $J(r)=1/r^2$, $E_N=\ln N+O(1)$ and $C_-=C_+=1$.

\textit{Existence of order}.---The role of the logarithm can be seen from an energy--entropy estimate due to Thouless~\cite{Thouless_1969lr}. At inverse temperature $\beta$, a reversed domain of length between $N$ and $2N$ that covers a given site can be placed in about $N^2$ ways, an entropy of about $2\ln N$, while its dimensionless cost is at least $4\beta E_N$. Their total Boltzmann weight, at most of order $N^2e^{-4\beta E_N}$, vanishes when $4\beta E_N-2\ln N\to\infty$. This estimate is not a proof, because a thermal fluctuation consists of many domains of all sizes whose energies do not add.

The estimate shows, however, that a diverging domain energy is not enough: it must keep growing at least logarithmically on all sufficiently large scales. If $E_N/\ln N$ dips to zero along some sequence of lengths, domains of those lengths are cheap at every temperature. Its conclusion is in fact exact, with the lower envelope $C_-$ as the deciding quantity.

\begin{theorem}[Existence of order]\label{thm:existence}
For $J$ as in Eq.~\eqref{eq:model}, the spontaneous magnetization $m_+(\beta)$ satisfies
\begin{equation}
\exists\,0<\beta<\infty:\ m_+(\beta)>0\iff C_->0 ,
\end{equation}
and when $C_->0$ the critical inverse temperature satisfies $0<\beta_c<\infty$.
\end{theorem}

We now outline the proof~\cite{supplement}. Necessity follows from a uniqueness theorem of Berbee~\cite{Berbee_1989uo}, which limits how strongly distant spins can affect the spins in a finite region. Applied to Eq.~\eqref{eq:model}, it states that if
\begin{equation}
\sum_{n\ge1}e^{-8\beta E_n}=\infty,
\label{eq:berbee}
\end{equation}
there is only one infinite-volume Gibbs measure: the equilibrium state of the infinite chain does not depend on the boundary conditions imposed far away. The states obtained with all spins up and with all spins down outside then coincide, and spin-flip symmetry gives $m_+=0$. The domain energy enters because $8\beta\sum_{r\ge k}J(r)$ bounds the change of the dimensionless energy of the two end spins of the interval $\{-n+1,\dots,n\}$ when all spins more than $k$ sites beyond them are changed arbitrarily, and these bounds sum to $\sum_{k\le n}8\beta\sum_{r\ge k}J(r)=8\beta E_n$. If $C_-=0$, there are lengths $N_\ell\to\infty$ with $E_{N_\ell}/\ln N_\ell\to0$, and since $E_n$ is nondecreasing,
\begin{equation}
	\sum_{n\le N_\ell}e^{-8\beta E_n}\ge N_\ell\,e^{-8\beta E_{N_\ell}}\xrightarrow[\ell\to\infty]{}\infty
\end{equation}
at every finite $\beta$, so Eq.~\eqref{eq:berbee} holds and $m_+=0$. For example, $J(r)=a/[r^2(\ln r)^\delta]$, $r\ge3$, $0<\delta<1$, has $E_N\sim a(\ln N)^{1-\delta}/(1-\delta)$, hence $C_-=0$ and no ordered phase, although its first moment and its domain energy both diverge. The bound $\beta_c>0$ follows from high-temperature uniqueness for $2\beta \Sigma_J<1$~\cite{Dobruschin_1968td}.

Sufficiency requires control of many domains at once. Reversing a finite set $A$ of spins costs $2B_J(A)$, where $B_J(A)=\sum_{x\in A,\,y\notin A}J(|x-y|)$ is the coupling across its boundary; an interval of $N$ sites has $B_J=2E_N$. A bond length $r$ crosses the boundary of a set at least $\min(r,L)$ times if the set contains an interval of $L$ sites. A set of $n$ sites made of $k$ runs therefore obeys
\begin{equation}
B_J(A)\ge2E_{\lceil n/k\rceil},
\label{eq:runs}
\end{equation}
while its $2k$ walls cost at least in proportion to $k$. Breaking a domain into fragments thus never makes it cheaper than its typical fragment, and balancing the two bounds gives the minimum cost of a defect of $n$ to $2n$ sites,
\begin{equation}
\liminf_{n\to\infty}\frac{\inf_{n\le|A|\le2n}B_J(A)}{\ln n}=2C_- ,
\label{eq:cut}
\end{equation}
when the distances carrying a bond have greatest common divisor one; otherwise the chain splits into decoupled sublattices, each covered by the same argument~\cite{supplement}. Thus the energy of a single interval bounds from below the energy of every defect.

To deduce order from Eq.~\eqref{eq:cut}, we use the Fortuin--Kasteleyn random-cluster representation, in which $m_+$ is the probability that the origin belongs to an infinite cluster~\cite{Fortuin_1972ot,Edwards_1988go,Aizenman_1988do}. The Ising cluster measure at inverse temperature $\beta$ dominates independent bond percolation in which a bond of length $r$ is open with probability $1-e^{-\beta J(r)}$~\cite{Edwards_1988go}, so it suffices to show that independent bonds percolate at large $\beta$. A cluster is cut off from the rest only if every bond across some cut is closed, which for a given cut has probability $e^{-\beta B_J}$; Eq.~\eqref{eq:cut} therefore bounds the probability of every cut. When the cut energies grow fast enough, $\sum_jE_{2^j}^{-1}<\infty$, a percolation theorem for weighted graphs~\cite{Hartarsky_2026wi} applies directly.

This condition fails when the domain energy grows only logarithmically. Whenever $\sum_jE_{2^j}^{-1}$ diverges, we instead build the infinite cluster scale by scale, adapting the block strategy of Refs.~\cite{Newman_1986od,Duminil-Copin_2024lr}. The block sizes are not taken from a geometric sequence but chosen from the coupling itself, at lengths where the bonds longer than a few blocks carry only a small share of the domain energy; such lengths recur without bound, since otherwise $E_N$ would grow as a power of $N$ and the sum would converge. Suppose that a cluster occupying most of each block of the current size fails to exist only with small probability. The blocks then act as sites joined by the longer bonds, and the next size is chosen so that these bonds join the blocks into a cluster occupying most of the larger block, again by the percolation estimate for finite graphs, while the bonds reserved for later sizes stay weak. At each step the part of the domain energy carried by bonds within a block at least doubles, the failure probability decays exponentially in this part, and the fraction of sites lost from the large cluster can be made summable. The construction starts from one sufficiently large initial block, made connected with high probability by a single finite increase of the inverse temperature~\cite{supplement}. The cluster densities therefore remain bounded away from zero while the failure probabilities tend to zero, so the origin belongs to an infinite cluster with positive probability, and the comparison with the Ising cluster measure gives $m_+>0$ at a finite inverse temperature. Choosing each size from the coupling accommodates arbitrary gaps between the interaction distances. The same random-cluster comparison extends the criterion to every finite-state Potts chain with the same $J$~\cite{Edwards_1988go,Grimmett_Book,supplement}.

\textit{Onset of order}.---\Cref*{thm:existence} leaves open how the magnetization appears at $\beta_c$: continuously from zero, as in most transitions, or with a jump. At finite temperature a long bond couples two distant regions only through their ordered parts, with an effective strength $m_+^2J(r)$, so a reversed region of length $N$ costs at most about $4\beta C_+m_+^2\ln N$ against the placement entropy $2\ln N$. If the magnetization were small, such regions would proliferate and destroy the order; this is avoided only if $2\beta C_+m_+^2\ge1$.

\begin{theorem}[Onset of order]\label{thm:onset}
Let $C_+<\infty$.

(i)~For every $0\le\beta<\infty$,
\begin{equation}
m_+(\beta)=0\qquad\text{or}\qquad 2\beta C_+\,m_+(\beta)^2\ge1 .
\label{eq:gap}
\end{equation}

(ii)~If in addition $C_->0$, then $0<\beta_c<\infty$ and
\begin{equation}
m_+(\beta_c)\ge(2\beta_cC_+)^{-1/2}>0 .
\label{eq:jump}
\end{equation}
\end{theorem}

\begin{figure}[t]
\centering
\includegraphics[width=\columnwidth]{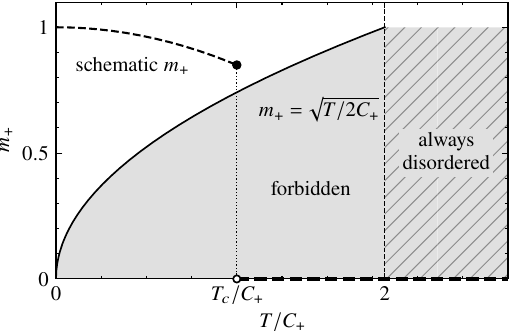}
\caption{Magnetization gap of \cref*{thm:onset}. Nonzero magnetizations below $m_+=\sqrt{T/(2C_+)}$ are forbidden (shaded), so every temperature $T\ge2C_+$ is always disordered (hatched). A schematic $m_+(T)$ (dashed) stays above the forbidden region and drops to zero at $T_c$, taking the upper value there (filled dot). The temperature $2C_+$ bounds $T_c$ from above and is not the transition temperature.}
\label{fig:gap}
\end{figure}

At the finite critical temperature a nonzero magnetization is thus bounded below by a strictly positive value, so the magnetization cannot appear continuously. Part~(i) does not use the existence of order. In terms of the temperature $T=1/\beta$, it states that $m_+=0$ or $m_+\ge\sqrt{T/(2C_+)}$ (Fig.~\ref{fig:gap}); since $m_+<1$ at finite temperature, the system must be disordered when $T\ge2C_+$. This is an upper bound on every ordered temperature, not the critical temperature itself.

The proof makes the picture of Thouless quantitative~\cite{Thouless_1969lr,Aizenman_1986do,Aizenman_1988do}. In the random-cluster representation, an open bond $\{x,y\}$ ties distant regions into the ordered cluster only if each end is joined to its surroundings without using the bond; we then call it a bridge. An end escapes its box of radius $R$ with probability $\rho_R$, which decreases to $m_+$ as $R\to\infty$. Exposing the two boxes one after the other gives
\begin{equation}
\Pr(\{x,y\}\text{ is a bridge})\le p_\beta(|x-y|)\bigl[\rho_R^2+o(1)\bigr],
\label{eq:bridge}
\end{equation}
where $p_\beta(r)=1-e^{-2\beta J(r)}$ is the bond probability and the correction vanishes as $|x-y|\to\infty$. This is the origin of the effective strength $m_+^2J(r)$.

To turn this estimate into a proof, we extend the renormalization of unbridged blocks developed for the inverse-square coupling~\cite{Duminil-Copin_2024lr}. Let $u(s)$ be the probability that a block of about $6s$ sites, with wired boundary, is not crossed by a path of bonds no longer than $s$. Inside a block of size $\lambda s$, call a sub-block of size $s$ unbridged if every long bond near it is closed and no bond spanning it is a bridge. Every crossing of the large block must also cross each unbridged sub-block, and each prevents the crossing with probability at least $u(s)$. The number of candidate sub-blocks is proportional to $\lambda$, giving, at every real scale $s\ge s_0$ and for all sufficiently large $\lambda$,
\begin{equation}
u(\lambda s)\ge c_1\lambda\,e^{-c_2\beta sF_s-2\beta\eta^2\mathcal M(s,\lambda)}\min\{u(s),\lambda^{-1}\},
\label{eq:recursion}
\end{equation}
where $0<c_1<1$ and $c_2>0$ are fixed numerical constants, $m_+<\eta<1$, $F_s=\sum_{r>s}J(r)$, and $\mathcal M(s,\lambda)=\sum_{s<r\le\lambda s}rJ(r)$ is the weight of the bonds that span a sub-block. The first exponent is the cost of closing the long bonds near a sub-block; the second is the cost of the spanning bonds, which carry the squared factor of Eq.~\eqref{eq:bridge}.

Along a geometric sequence of scales, Eq.~\eqref{eq:recursion} closes when $sF_s$ and $\mathcal M(s,\lambda)/\ln\lambda$ are bounded at every scale, as for the inverse-square coupling~\cite{Duminil-Copin_2024lr}. For general couplings they are not: there are chains with $C_+<\infty$ whose $sF_s$ is unbounded~\cite{supplement}. We therefore use Eq.~\eqref{eq:recursion} at every real scale and average its logarithm over $\ln s$. Both costs then reduce to the domain energy. With $E(x)=\sum_r\min(r,x)J(r)$ at real $x$,
\begin{equation}
\begin{aligned}
\int_{s_0}^{X}sF_s\,\frac{ds}{s}&=E(X)-E(s_0),\\
\int_{s_0}^{X}\mathcal M(s,\lambda)\,\frac{ds}{s}&\le(\ln\lambda)\,E(\lambda X),
\end{aligned}
\label{eq:average}
\end{equation}
so each bond contributes only over the scales where it acts.

The cutoff $\min\{u(s),\lambda^{-1}\}$ needs separate control. Assume $m_+>0$ and write $q(s)=-\ln u(s)$. In an ordered state a growing block connects, with probability tending to one, to sites farther than a fixed multiple of its size; closing the long bonds around the block and blocking short-bond crossings would prevent this, which gives $2q(s)+2c_2\beta sF_s\to\infty$~\cite{supplement}. Scales with $q(s)<\ln\lambda$ must therefore have large $sF_s$, and since Eq.~\eqref{eq:average} bounds the logarithmic average of $sF_s$, they occupy a vanishing fraction of the logarithmic scale range. The cutoff therefore does not change the logarithmic average of Eq.~\eqref{eq:recursion}.

Taking minus the logarithm of Eq.~\eqref{eq:recursion} and integrating over $ds/s$, the difference $q(\lambda s)-q(s)$ leaves the integral over $[X,\lambda X]$ minus the fixed integral over $[s_0,\lambda s_0]$. The first is nonnegative because $q\ge0$, so the difference has a lower bound independent of $X$. Dividing by $\ln X$, neglecting the cutoff and using Eq.~\eqref{eq:average}, and letting $X\to\infty$ gives
\begin{equation}
0\le-\bigl(1-2\beta\eta^2C_+\bigr)\ln\lambda+\ln(1/c_1)+c_2\beta C_+ .
\end{equation}
If $m_+>0$ and $2\beta C_+m_+^2<1$, we may take $\eta>m_+$ with $2\beta C_+\eta^2<1$, and a large $\lambda$ makes the right side negative. This contradiction proves part~(i). The cost of the long bonds near a sub-block is fixed and is divided by $\ln\lambda$, while the spanning cost grows with $\ln\lambda$ and carries $\eta^2$; this is why a small nonzero magnetization cannot be sustained.

Part~(ii) then follows. \Cref*{thm:existence} supplies an ordered temperature, every $\beta>\beta_c$ is ordered because $m_+$ is nondecreasing in $\beta$, and the right-continuity of $m_+$ in $\beta$~\cite{Aizenman_1988do} carries the bound of part~(i) to $\beta_c$.

\textit{Scope and examples}.---Earlier jump theorems assume a bound on the interaction at every distance, a finite $\limsup_{r\to\infty}r^2J(r)$~\cite{Aizenman_1988do}. The hypothesis $C_+<\infty$ on the domain energy is never stronger. If $J(r)\le L/r^2$ for $r>r_0$, bonds shorter than $N$ contribute at most $L\ln N+O(1)$ to $E_N$ and longer ones at most $L$; letting $L$ decrease to $\limsup_{r\to\infty}r^2J(r)$ gives
\begin{equation}
C_+\le\limsup_{r\to\infty}r^2J(r)
\end{equation}
for every summable $J$. For $J(r)\sim c/r^2$, $C_-=C_+=c$, and \cref*{thm:onset} reproduces exactly the known alternative $2\beta c\,m_+^2\ge1$ and jump $m_+(\beta_c)\ge(2\beta_cc)^{-1/2}$, in our normalization, where the coupling of Ref.~\cite{Aizenman_1988do} is $2J$; this chain orders by the theorem of Fr\"ohlich and Spencer~\cite{Frohlich_1982tp}.

The inequality can be strict. As an example, consider a chain whose bonds exist only at the distances $1,2,4,8,\ldots$, with $J_a(2^k)=a\ln2\;2^{-k}$ for $k\ge0$ and amplitude $a>0$; we call it the dyadic chain. Its domain energy obeys $E_{2N}-E_N=a\ln2$ exactly: doubling the length of a domain always adds the same energy, so the domain energy grows logarithmically, at the marginal rate set by the placement entropy. Hence $E_N=a\ln N+O(1)$ and $C_-=C_+=a$, so the chain orders, $\beta_c>1/(2a)$, and its magnetization jumps, $m_+(\beta_c)\ge[2a\beta_c]^{-1/2}$. Yet its bonds occupy a vanishing fraction of distances and $r^2J_a(r)=a\ln2\;2^k$ is unbounded on its support, so the class $0<C_-\le C_+<\infty$ is strictly larger than the pointwise one. Couplings with irregular scales give further examples~\cite{supplement}.

The growth rates $C_-$ and $C_+$ do not, however, fix the critical temperature or the size of the jump. The couplings $J_L(r)=ar^{-2}$ for $r\ge L$ and zero below all have $C_-=C_+=a$, yet their critical temperature is at most $2a/(L-1)$ by high-temperature uniqueness and vanishes as $L\to\infty$. Spreading the bonds of $J$ onto the multiples of an integer $d\ge2$ splits the chain into $d$ decoupled copies, leaving $m_+$ and $\beta_c$ unchanged while multiplying $C_+$ by $d$, so the bound~\eqref{eq:jump} can be off by any factor~\cite{supplement}.

\textit{Discussion}.---For every summable ferromagnetic Ising chain, the domain energy, measured against the logarithm of the domain length, decides whether an ordered phase exists, and a logarithmic upper bound on it forces the magnetization to jump at the critical point. The discontinuity predicted by Thouless is thus a consequence of the logarithmic growth of the domain energy rather than of the inverse-square form of the coupling.

Besides the localization transition of the Ohmic spin-boson model~\cite{Chakravarty_1982qf,Spohn1_1985qt,Leggett_1987do,Spohn_1989gs,Chakravarty_1995dd,Hasler_2022cb,Betz_2025ot}, discontinuities of this type recur in mixed-order transitions~\cite{Bar_2014mo,Bar_2014mo-1,Bassler_2015et,Mukamel_2023mo}, and effective long-range couplings along a line make the transition of book Ising models discontinuous~\cite{Duminil-Copin_2023lr}.

Whether the bound~\eqref{eq:jump} is saturated for chains whose bonds generate the whole lattice, for the dyadic chain, or for $1/r^2$ itself, where Monte Carlo data indicate a universal jump of the scaled order parameter~\cite{Luijten_2001ci}, remains open, as it was in Ref.~\cite{Aizenman_1988do}. Further questions concern the transition when the domain energy has no logarithmic upper bound, $C_->0$ with $C_+=\infty$, and the critical behavior of the sparse and irregular chains covered here.

\textit{Acknowledgments}.---Z.~Y. and K.~C. acknowledge support from the Strategic Priority Research Program of the Chinese Academy of Sciences under Grant No.~XDB1680102. Z.\ Y.\ also acknowledges support from the National Natural Science Foundation of China under Grant No.~12247101. We acknowledge Gewu SciLab for providing the collaborative research environment in which this project was developed.

OpenAI GPT-6 Astra, Anthropic Claude Opus 5.5, and Claude Fable 5.1 were used throughout this work: to develop and check the arguments, to review the proofs and their hypotheses, to search and verify the literature, and to draft, organize and typeset the manuscript and prepare its figures. The authors take full responsibility for the content of this paper.

\textit{Data availability}.---No data were created or analyzed in this study.

\clearpage
\onecolumngrid
\setcounter{secnumdepth}{3}
\setcounter{equation}{0}
\setcounter{figure}{0}
\setcounter{table}{0}
\setcounter{section}{0}
\renewcommand{\theequation}{S\arabic{equation}}
\renewcommand{\thefigure}{S\arabic{figure}}
\renewcommand{\thetable}{S\arabic{table}}
\renewcommand{\thesection}{\Roman{section}}
\renewcommand{\theHequation}{S\arabic{equation}}
\renewcommand{\theHfigure}{S\arabic{figure}}
\renewcommand{\theHtable}{S\arabic{table}}
\renewcommand{\theHsection}{S\Roman{section}}

\begin{center}
{\large\bfseries Supplemental Material for ``Onset of Ferromagnetic Order in Ising Chains''}
\end{center}
\vspace{1em}

In this Supplemental Material, we provide detailed derivations of the following
topics: the uniqueness and cut estimates behind the existence criterion; the
scale-by-scale percolation construction that proves order whenever $C_->0$; the
extension of the criterion to Potts chains; the pressure, magnetic-response and
finite-range bounds on $\beta_c$; the proof of both parts of Theorem~2 of the main text,
the magnetization gap and the jump at $\beta_c$; the dyadic, pulse and dilated chains; and the
inequality $C_+\le\limsup_{r\to\infty}r^2J(r)$.

\section{Uniqueness when the domain energy grows sublogarithmically}
\label{sec:sm-uniqueness}

We follow the notation and hypotheses of the main text throughout, for general
summable $J\ge0$. All constants used below are deliberately nonoptimal. In the
quantitative construction we fix an energy unit, so exponential tail bounds are
written for numerical values of $J$; restoring an arbitrary reference energy
changes neither the sign of $C_-$ nor the existence statement. The argument is
entirely analytic.

Berbee's two-sided Gibbs criterion~\cite{Berbee_1989uo} is stated
in terms of uniform shell variations $v_k$. For the pair interaction of the main
text, symmetric interval shells give $v_k\le8\beta\sum_{r\ge k}J(r)$ and
consequently $\sum_{k=1}^n v_k\le8\beta E_n$. If $C_-=0$, choose
$N_\ell\to\infty$ with $E_{N_\ell}/\ln N_\ell\to0$. Then
\begin{equation}
 \sum_{n\le N_\ell}e^{-8\beta E_n}
 \ge N_\ell e^{-8\beta E_{N_\ell}}\longrightarrow\infty .
 \label{eq:sm-berbee}
\end{equation}
The criterion then gives uniqueness for every finite $\beta$, and spin-flip
symmetry gives zero magnetization. This is a Gibbs-specification argument on
$\ZZ$, not an inference from uniqueness of a one-sided chain. In particular,
$r^2J(r)\to0$ implies $E_N=o(\ln N)$: for each $\eta>0$ the tail is bounded by
$\eta/r^2$, giving $S_N\le O_\eta(1)+\eta\ln N$ and $NF_N\le\eta$.

For nonzero $J$, the usual single-site contraction bound $2\beta \Sigma_J<1$ gives a
positive high-temperature uniqueness interval. Thus existence of order at a
finite inverse temperature implies $0<\beta_c<\infty$.

\section{Every finite reversed set faces the interval barrier}
\label{sec:sm-cut}

For a finite nonempty $A\subset\ZZ$ of size $n$, put $d_r=n-|A\cap(A+r)|$. In
the graph whose edges have length $r$, $d_r$ is the number of finite components
of $A$. If $A$ contains an interval of length $L$, at least $\min(r,L)$ residue
classes modulo $r$ are occupied, so $d_r\ge\min(r,L)$. Since
$B_J(A)=2\sum_rJ(r)d_r$, a set with $k$ consecutive runs has
\begin{equation}
 B_J(A)\ge 2E_{\lceil n/k\rceil}.
 \label{eq:sm-runs}
\end{equation}

Suppose that the support of $J$ has greatest common divisor one. Choose a word of
supported signed steps $s_1,\ldots,s_\ell$ summing to one, and let
$j_*=\min_iJ(|s_i|)>0$. Translate the resulting path over all nearest-neighbor
edges. Each of the $2k$ nearest-neighbor boundary edges of $A$ has a crossing
somewhere along its path. Each supported crossing is charged at most $\ell$ times,
hence $2k\le(\ell/j_*)B_J(A)$. Set $c_0=2j_*/\ell$.

Write $\psi_J(n)=\inf\{B_J(A):n\le|A|\le2n\}$. For any $\kappa\in(0,1)$,
splitting at $k=n^{1-\kappa}$ between Eq.~\eqref{eq:sm-runs} and the
fragmentation bound gives
\begin{equation}
 \min\{c_0n^{1-\kappa},\,2E_{\lfloor n^\kappa\rfloor}\}
 \le\psi_J(n)\le2E_n ,
 \label{eq:sm-psibounds}
\end{equation}
the upper bound being the cost of an interval. If $C_-<\infty$, the lower
$\liminf$ of $\psi_J(n)/\ln n$ is at least $2\kappa C_-$; letting
$\kappa\uparrow1$ gives
\begin{equation}
 \liminf_{n\to\infty}\frac{\psi_J(n)}{\ln n}=2C_- .
 \label{eq:sm-cut}
\end{equation}
If $C_-=\infty$, the same bound gives infinity. Equation~\eqref{eq:sm-cut} is the
cut bound quoted in the main text.

For a support of greatest common divisor $d$, use the contraction
$\bar J_d(r)=J(dr)$. Then $E_n(\bar J_d)=E_{dn}(J)/d$, so the coefficient of
$\bar J_d$ is $C_-/d$. Every set of $n$ vertices has at least $n/d$ vertices in
one residue class, which gives the lower $\liminf$ $2C_-/d$; an interval within a
single class gives the matching upper bound.

\section{The two branches of the existence proof}
\label{sec:sm-branches}

\subsection{A summable reciprocal profile percolates directly}
\label{sec:sm-summable}

Write $\mathcal R=\sum_{j\ge0}E_{2^j}^{-1}$. If $\mathcal R<\infty$, the lower bound
in Eq.~\eqref{eq:sm-psibounds} with $\kappa=1/2$ gives
\begin{equation}
 \sum_{j\ge0}\psi_J(2^j)^{-1}<4/c_0+\mathcal R .
 \label{eq:sm-summable}
\end{equation}
The weighted-percolation theorem of Ref.~\cite{Hartarsky_2026wi}, applied at a sufficiently large multiplier,
gives an infinite cluster with positive probability. The random-cluster
comparison of Sec.~\ref{sec:sm-passage} then gives Ising order. The remainder of
the construction, Secs.~\ref{sec:sm-giant}--\ref{sec:sm-renewal}, handles the
complementary branch $\mathcal R=\infty$. The all-cut logarithmic estimate
\eqref{eq:sm-cut} alone does not make $\mathcal R$ converge.

\subsection{The tail cannot be too small, and admissible scales recur}
\label{sec:sm-scalar}

Assume $C_->0$. Then $S_R\to\infty$ and $F_R>0$ at every finite $R$. For all
$R,N$, $E_N\le S_R+NF_R$. Take $N=\lfloor1/F_R\rfloor$ and use $E_N\ge c\ln N$
for a fixed $c>0$ and all large $N$. There exist $C>0$ and $R_*$ such that
\begin{equation}
 F_R\ge e^{-CS_R}\quad(R\ge R_*).
 \label{eq:sm-tailbound}
\end{equation}

In the branch $\mathcal R=\infty$, for $C_0=10^6$ there are arbitrarily large even
$K$ with
\begin{equation}
 3KF_{3K}\le S_{3K}/(16C_0).
 \label{eq:sm-admissible}
\end{equation}
Otherwise eventually $NF_N>S_N/(32C_0)$. The function
$E(x)=\sum_r\min(r,x)J(r)$ is linear between integers, with
$E'(x)=F_{\lfloor x\rfloor}$ almost everywhere. Then
$xE'(x)\ge E(x)/(32C_0+1)$ eventually, forcing polynomial growth of $E$ and
contradicting $\mathcal R=\infty$. Thus there are arbitrarily large integers with
the reversed inequality. Rounding each up to a multiple of six changes it by at
most a factor two and proves Eq.~\eqref{eq:sm-admissible}. We call such $K$
admissible.

\section{A giant component in a finite interval}
\label{sec:sm-giant}

\subsection{The restricted cut bound}
\label{sec:sm-restricted}

For a finite graph $G=(V,E)$ with bond rates $w_e\ge0$, define $\psi_w(n)$ as
the minimum cut capacity over sets $S$ with $n\le|S|\le2n$ and $|S|\le|V|/2$.
Its reciprocal budget is $\sum_{j\ge0:\,2^j\le|V|/2}1/\psi_w(2^j)$. We use
Theorem~1.1 of Ref.~\cite{Hartarsky_2026wi}: independent bonds of probabilities
$1-e^{-w_e}$ and budget at most $10^{-4}$ satisfy
\begin{equation}
 \PP(x\not\leftrightarrow y)\le e^{-3\Phi(x,y)/4},
 \label{eq:sm-hst}
\end{equation}
where $\Phi(x,y)$ is the minimum capacity of a cut separating $x$ from $y$.

The following restricted form is a consequence of Lemma~2.4 and Eq.~(19) of
Ref.~\cite{Hartarsky_2026wi}. Write $\mathcal C(x)$ for the open cluster of $x$.
If every cut with both sides at least $n_0$ has capacity at least $P(n_0)$, then
\begin{equation}
 \PP\bigl(x\not\leftrightarrow y,\ |\mathcal C(x)|\ge n_0,\ |\mathcal C(y)|\ge n_0\bigr)
 \le e^{-3P(n_0)/4}.
 \label{eq:sm-restrictedcut}
\end{equation}
To verify this restriction, let $\mathfrak S_{n_0}(x,y)$ be the deterministic
family of minimal separating cutsets whose two sides both have at least $n_0$
vertices, and write $w(\pi)=\sum_{e\in\pi}w_e$ for the capacity of a cutset
$\pi$. The cited equation gives, for the random output $\Pi$ of the cutset
exploration,
\begin{equation}
 \PP_w(\pi\text{ is cohesive})
 \le e^{-(1-\varepsilon)w(\pi)}\Pr(\Pi=\pi),
 \qquad \varepsilon\le1/4.
 \label{eq:sm-cohesive}
\end{equation}
On the joint event in Eq.~\eqref{eq:sm-restrictedcut}, the cited lemma supplies
a closed cohesive cutset. Its sides contain the entire corresponding root
clusters, so it belongs to $\mathfrak S_{n_0}(x,y)$. Therefore
\begin{equation}
 \PP_w\bigl(x\not\leftrightarrow y,\ |\mathcal C(x)|,|\mathcal C(y)|\ge n_0\bigr)
 \le\sum_{\pi\in\mathfrak S_{n_0}(x,y)}
 e^{-3w(\pi)/4}\Pr(\Pi=\pi)
 \le e^{-3P(n_0)/4}.
\end{equation}
The final step uses $w(\pi)\ge P(n_0)$ and the fact that the exploration
probabilities sum to at most one on this restricted family. The estimate is
unconditional; no independence of the two cluster-size events is required. The
derivation below checks the budget again after terminal wiring and ghost
addition, before applying the same external theorem.

\subsection{Rates whose budget is small produce a giant component}
\label{sec:sm-finitegiant}

Let $R\ge2$, $N=4R^2$, and $V=\{0,\ldots,N-1\}$. Nearest-neighbor bonds have
rate $b$, and length-$m$ bonds have rate $a_m\ge0$, $2\le m\le R$. Write
\begin{equation}
 A=\sum_m a_m,\quad M=\sum_m ma_m,\quad U(x)=\sum_m\min(m,x)a_m,
 \qquad P(s)=b/2+U(\sqrt{s})/4 .
\end{equation}
For $0<\delta\le1/32$, assume
\begin{equation}
 b\ge2A,\quad b\ge8192\ln(1/\delta),\qquad
 \frac1b+\sum_{\substack{j\ge0\\2^j\le N/2}}\frac1{P(2^j)}
 \le10^{-4}.
 \label{eq:sm-giantbudget}
\end{equation}
We show that an internal component then has at least $(1-\delta)N$ vertices
except with probability at most $(2+\delta^{-2})e^{-\delta M/128}$.

For $\varnothing\ne S\subset V$, $s=|S|\le N/2$, let $c_m$ count internal
length-$m$ edges crossing the boundary of $S$. We first show
\begin{equation}
 c_1+c_m\ge\tfrac14\min(\sqrt{s},m).
 \label{eq:sm-finitecut}
\end{equation}
For $s\le16$ use $c_1\ge1$. Otherwise either $c_1\ge\sqrt{s}/4$ or $S$ has at
most $c_1+1\le\sqrt{s}/2$ runs and contains a run of length $L\ge2\sqrt{s}$.
That run occupies at least $\min(m,L)$ residue classes modulo $m$. A non-full
occupied residue chain contributes a crossing. The number $f$ of full chains is
at most $s/\lfloor N/m\rfloor$. If $2\sqrt{s}\ge m$, then $f\le2m/3$, so
$c_m\ge m/3$. If $2\sqrt{s}<m$, then $f\le s/(4m)<\sqrt{s}/8$, so
$c_m\ge2\sqrt{s}-f$. This proves Eq.~\eqref{eq:sm-finitecut}. Therefore every
cut has capacity at least $b$ and, since $b\ge2A$, at least $P(s)$.

Wire any nonempty list of terminals to $x$ and add a ghost vertex $g$, joined
with rate $t_g/N$ to each original vertex, where $t_g\ge b$. A quotient cut whose
smaller side has $s\ge2$ vertices leaves at least $s-1$ original vertices on
each side after undoing the wiring and removing $g$. Its capacity is at least
$P(s-1)$. Every nontrivial quotient cut has capacity at least $b$: either both
sides contain original vertices and a nearest-neighbor edge crosses, or one side
contains only the ghost and has capacity $t_g\ge b$. If the quotient has $n'$
vertices, its profile consequently obeys
\begin{equation}
 \frac1{\psi_{\rm quot}(1)}
 +\sum_{\substack{j\ge1\\2^j\le n'/2}}
       \frac1{\psi_{\rm quot}(2^j)}
 \le\frac1b+
   \sum_{\substack{j\ge1\\2^j\le n'/2}}\frac1{P(2^{j-1})}
 \le\frac1b+\sum_{\substack{i\ge0\\2^i\le N/2}}\frac1{P(2^i)}.
\end{equation}
Here $n'\le N+1$, and $P$ is nondecreasing. Thus terminal wiring and the ghost
preserve the budget required for the cited theorem.

Two independent uniform roots remove a volume factor from the restricted cut
bound. If two distinct components have size at least $n_0$, the probability that
the roots lie in them is at least $(n_0/N)^2$. Hence
\begin{equation}
 \PP(\text{two components of size at least }n_0)
 \le(N/n_0)^2e^{-3P(n_0)/4}.
 \label{eq:sm-twolarge}
\end{equation}
Also, in the absence of a $(1-\delta)N$ component, the conditional probability
of root disconnection is $1-\sum_{\mathcal C}(|\mathcal C|/N)^2\ge\delta$, the
sum running over components. Thus the failure probability is at most
$\delta^{-1}e^{-3b/4}$. This proves the claim when $\delta M<32b$.

In the remaining case put $k=\lfloor\delta M/(16b)\rfloor\ge2$ and $t_g=4bk$.
Draw $1024k$ uniform terminals. Call a terminal list good if every union of at
most $k$ intervals containing it has more than $\delta N/2$ vertices. Every cut
separating such a list from the ghost has capacity at least $bk$. Indeed, an
original set $Q$ containing the list has ghost capacity $>2bk$ if
$|Q|>N/2$. Otherwise either $c_1(Q)\ge k$, or $Q$ has at most $k$ runs and
size $>\delta N/2$. Since $M\le AR$ and $b\ge2A$, $k\le\delta R/32$; its longest
run then has length at least $\delta N/(2k)\ge64R$. All residue classes for
$m\le R$ are occupied and $c_m(Q)\ge m/3$, giving capacity at least
$M/3\ge bk$.

Let $\mathcal E_\delta$ be the event that a union $S$ of components has
$\delta N\le|S|\le2\delta N$. Choose $S$ deterministically from the bond
configuration. All terminals fall in it with probability at least
$\delta^{1024k}$. Conditional on this, the sampling estimate of
Sec.~\ref{sec:sm-sampling} makes the list good with probability at least $1/2$.
Independently, no ghost edge touches $S$ with probability at least
$e^{-8\delta bk}$. These events disconnect $x$ from $g$, so
\begin{align}
 \PP(\mathcal E_\delta)
 &\le2\exp\{- (3/4-8\delta)bk+1024k\ln(1/\delta)\}\nonumber\\
 &\le2e^{-bk/4}\le2e^{-\delta M/128}.
 \label{eq:sm-smallcomponents}
\end{align}
If no giant exists, either two components have size at least $\delta N$ or
$\mathcal E_\delta$ occurs, by greedily collecting smaller components. At
$n_0=\lceil\delta N\rceil$, $P(n_0)\ge b/2+\sqrt{\delta}M/2$, since $m\le R$.
The bound \eqref{eq:sm-twolarge} is then at most $\delta^{-2}e^{-\delta M/128}$.
Adding Eq.~\eqref{eq:sm-smallcomponents} gives the claimed failure probability.
No domination of the nonmonotone event in Eq.~\eqref{eq:sm-twolarge} is used.

\subsection{Random points pin down a majority by intervals}
\label{sec:sm-sampling}

For any finite ordered set $S$, draw $r=1024k$ independent uniform points of $S$
with replacement. We show that, with probability at least $1/2$, every union of
at most $k$ intervals containing the sample contains more than $|S|/2$ points of
$S$.

Let $X,Y$ be independent samples of size $r$. If $X$ is bad, choose a witnessing
interval union of measure at most $1/2$. With probability at least $1/2$, at
least $r/4$ of the $Y$ points are outside it (the failure probability is at most
$4/r$). Condition on the pooled $2r$ labeled points and randomly assign $r$
labels to $X$. There are at most $\sum_{j\le2k}\binom{2r+1}{j}\le(3er/k)^{2k}$
traces of interval unions. A trace containing at most $7r/4$ pooled labels
contains all $X$ labels with probability at most $(7/8)^r\le e^{-r/8}$. A union
bound gives a bad-$X$ probability at most
$2(3er/k)^{2k}e^{-r/8}\le2e^{-108k}<1/2$. The argument allows repeated sample
points.

\section{Transfer of density from blocks to their parent}
\label{sec:sm-transfer}

An internal event uses only bonds whose two endpoints are in its interval. At an
even scale $K$, suppose a component of density $7/8$ fails with probability at
most $u_w$, and a component of density $\theta\ge24/25$ fails with probability
at most $u_s$, uniformly over translates. Put $y_s=-\ln u_s$; a positive upper
bound may replace $u_s=0$. For fresh lengths $j>3K$, define the coarse rates
\begin{equation}
 a_m=\frac{\beta K}{4}
       \sum_{\substack{j>3K\\\lfloor j/K+1/2\rfloor=m}}J(j),
 \quad A_R=\sum_{m\le R}a_m,\quad
 M_R=\sum_{m\le R}ma_m,\quad
 U_R(x)=\sum_{m\le R}\min(m,x)a_m .
 \label{eq:sm-coarserates}
\end{equation}
Only $m\ge3$ can occur, and zero rates may be inserted at $m=2$. The weak and
strong events refer to the same internal bond configuration. A core means the
entire component meeting the specified density threshold; such a component is
unique because the threshold exceeds one half.

Let $R\ge3$, $N=4R^2$, $B=NK$, and $0<\delta\le1/32$, and assume
\begin{align}
 3u_w&\le e^{-2b}/4,&
 b&\ge16A_R,& b&\ge131072\ln(1/\delta),\nonumber\\
 \frac{16}{b}
 +8\sum_{\substack{j\ge0\\2^j\le2R^2}}
 \frac1{b/4+U_R(\sqrt{2^j})}&\le10^{-4},&
 y_s&\ge2\ln(e/\delta),&
 M_R&\le128Ny_s .
 \label{eq:sm-transferbudget}
\end{align}
We show that an internal parent component of density $\theta(1-2\delta)$ in an
interval of length $B$ then fails with probability at most
$(3+\delta^{-2})e^{-\delta M_R/256}$, and an internal parent component of density
$7/8$ fails with probability at most $1027e^{-M_R/8192}$.

\subsection{Good seeds dominate independent sites}
\label{sec:sm-seeds}

Partition the line into primary intervals $I_i=[iK,(i+1)K)\cap\ZZ$ and bridges
$D_i=I_i+K/2$. Declare a seed $I_i$ good when the weak event holds in
$I_i,D_{i-1},D_i$. Its failure probability is at most $\varepsilon=3u_w$. Good
cores in overlapping windows intersect: their overlap contains at least
$(2(7/8)-3/2)K=K/4$ common core sites. Thus adjacent good seeds connect through
their common bridge. The seed indicators are 1-dependent: collections separated
by at least one seed depend on disjoint internal edge sets. The product-domination
bound~\cite{Liggett_1997db,Temmel_2012sm} gives independent retained seeds with
retention at least $(1-\sqrt\varepsilon)^2\ge1-2\sqrt\varepsilon\ge1-e^{-b}$.
The retained indicators are auxiliary independent variables coupled below the
original good-seed indicators.

\subsection{Fresh long bonds join retained seeds independently}
\label{sec:sm-freshbonds}

Let $\Xi_i$ be the weak core in a retained seed. For a microscopic length
assigned to a coarse displacement $m$, the overlap of the two shifted primary
intervals has at least $K/2$ sites. Removing their two core deficits leaves at
least $K/4$ available pairs. Let $\mathcal F$ contain the old edges and the
auxiliary randomness used to retain seeds. For each pair of retained seeds at
displacement $m$, its available fresh edges have total rate $L_e\ge a_m$. For
$a_m>0$, conditionally on $\mathcal F$, thin their connection indicator using an
independent uniform variable with retention probability
$(1-e^{-a_m})/(1-e^{-L_e})$. If an endpoint is absent, instead supply an
auxiliary Bernoulli bond of probability $1-e^{-a_m}$; it will be deleted in the
site model. Zero-rate bonds are always closed. Different coarse pairs use
disjoint microscopic bonds and independent auxiliaries. Thus their conditional
joint law is the same product law for every $\mathcal F$, making the potential
bonds independent of the retained sites. This does not require the retained
sites to remain independent conditional on the old edges. Automatic
adjacent-seed connections receive independent potential bonds of rate $b/8$,
with the same convention for absent endpoints. These constructions provide paths
internal to the parent whenever both seeds lie in it. The outer bridges are only
certification devices; connections between parent seeds use internal bridges.

\subsection{Absent sites cost at most half of each bond rate}
\label{sec:sm-sitebond}

We record the site-to-bond comparison used here. Consider independent sites of
failure probability at most $e^{-b}$ and independent potential bonds,
independent also of the sites, of rates $\lambda_e\le A'$ and incident total
rate at most $4A'$, and assume $A'\ge1$ and $b\ge8A'$. Delete bonds touching
absent sites. This model dominates independent bonds of rates $\lambda_e/2$.

To see this, reveal surviving edges sequentially and fix a positive-probability
history $\mathcal H$. If a revealed open edge touches $v$, its absence
probability is zero. Otherwise let $L_v$ be the sum of rates of the revealed
edges incident to $v$. The nonincident observations are independent of the prior
state of $v$. Given those observations and all other sites, the likelihood of the
incident closed observations is one when $v$ is absent and at least $e^{-L_v}$
when it is present. Averaging over the other sites therefore bounds the
whole-history likelihood ratio by $e^{L_v}$. With prior absence probability
$q_v\le e^{-b}$, Bayes' rule gives
\begin{equation}
 \frac{\PP(v\text{ absent}\mid\mathcal H)}
      {\PP(v\text{ present}\mid\mathcal H)}
 \le\frac{q_v}{1-q_v}e^{L_v}
 \le\frac{e^{-b+4A'}}{1-e^{-b}}
 \le2e^{-4A'}=:\eta .
\end{equation}
The conditional absence probability is at most these odds. The next potential
bond is independent of the history and the sites. A union bound on its two
endpoints, without conditional independence of their states, gives opening
probability at least $(1-2\eta)(1-e^{-\lambda_e})\ge1-e^{-\lambda_e/2}$.
Sequential uniform coupling proves domination. The configuration-independent
thinning of Sec.~\ref{sec:sm-freshbonds} supplies the required site--bond
independence. Here take $A'=b/8$. The total incident rate is
$b/4+2A_R\le3b/8\le4A'$, so the comparison gives nearest-neighbor rate $b/16$
and other rates $a_m/2$.

\subsection{The coarse giant carries the strong cores}
\label{sec:sm-coarsegiant}

Apply the giant estimate of Sec.~\ref{sec:sm-finitegiant} to this bond graph. Its
profile is $(b/4+U_R(\sqrt{s}))/8$, and Eq.~\eqref{eq:sm-transferbudget} is
precisely a sufficient reciprocal budget. The coarse giant fails with probability
at most $(2+\delta^{-2})e^{-\delta M_R/256}$.

The strong events in the $N$ disjoint primary intervals are independent. If $X$
counts their failures, a binomial bound gives, for $d>0$,
\begin{equation}
 \PP(X>dN)\le
 \exp\{-dN[y_s-\ln(e/d)]\}.
\end{equation}
With $d=\delta$, this is at most $e^{-\delta Ny_s/2}$, and hence at most
$e^{-\delta M_R/256}$. When the coarse giant misses at most $\delta N$ seeds and
$X\le\delta N$, at least $(1-2\delta)N$ seeds in it contain strong cores. A
strong and a weak core in the same primary interval coincide, because both occupy
more than half the interval. Their union belongs to one parent component of the
required density. A union bound, which does not assume independence of $X$ and
the coarse giant, proves the strong estimate. Repeating the argument with
$d=\delta_w=1/32$ gives density $\theta(15/16)\ge0.9>7/8$ and failure
$(3+32^2)e^{-M_R/8192}$, which is the weak estimate.

\section{Selecting the next interaction scale}
\label{sec:sm-selection}

Throughout this section $a_m\ge0$, $m\ge2$, has finite positive mass $A$ and
infinite first moment. For real $x\ge1$ define
\begin{equation}
 U(x)=\sum_m\min(m,x)a_m,\quad
 M(x)=\sum_{m\le x}ma_m,\quad G(x)=\sum_{m>x}a_m,\quad g(x)=xG(x).
\end{equation}
Thus $U=M+g$ and $U'=G$ almost everywhere. Write
$U_R(x)=\sum_{m\le R}\min(m,x)a_m$. For $\nu\ge A$ put
\begin{equation}
 \mathcal Q_\nu(R)=\sum_{\substack{j\ge0\\2^j\le2R^2}}
       [\nu+U_R(\sqrt{2^j})]^{-1}.
 \label{eq:sm-budgetQ}
\end{equation}
All dyadic sums below start at $j=0$.

\subsection{A prefix reaching a prescribed moment at small budget}
\label{sec:sm-prescribed}

Given $Z\ge\nu$, let $r_*$ be the first integer with $M(r_*)\ge Z$ and let
$\Omega_*=\sum_{m\ge r_*}a_m$. We show that there is a finite prefix $R_a$ with
$M(R_a)\ge Z$ and
\begin{equation}
 \mathcal Q_\nu(R_a)\le40[1+\ln(\nu/\Omega_*)]/\nu ,
 \label{eq:sm-prescribed}
\end{equation}
and that, for the microscopic coarse rates~\eqref{eq:sm-coarserates}, if
Eq.~\eqref{eq:sm-tailbound} holds for $d\ge3K$, then
\begin{equation}
 \ln(\nu/\Omega_*)\le
 \ln\frac{4\nu}{\beta K}+C(S_{3K}+8Z/\beta).
 \label{eq:sm-microprescribed}
\end{equation}

First consider a temporary sequence $\widetilde a_m$ of mass
$0<\Omega=\sum_m\widetilde a_m\le\nu$. In this paragraph all prefix profiles are
formed from this sequence. Choose $\xi$ uniformly from $[\Omega/4,\Omega/2]$ and
define $R(\xi)=\min\{R:\sum_{m\le R}\widetilde a_m\ge\xi\}$ and
$\Omega_R=\sum_{m\le R}\widetilde a_m$. For $x<R(\xi)$, write
$A_x=\sum_{m\le x}\widetilde a_m$ and $S_x=\sum_{m\le x}m\widetilde a_m$. Then
$U_{R(\xi)}(x)\ge S_x+x(\xi-A_x)$. Average
$I(\xi)=\int_1^{R(\xi)}[x(\nu+U_{R(\xi)}(x))]^{-1}\,dx$. Where
$A_x<\Omega/8$, its contribution is at most $\nu^{-1}\ln(1+8\nu/\Omega)$. Where
$\Omega/8\le A_x<\Omega/2$, integrate in $\xi$ first and extend the lower
endpoint to $A_x$. Dropping $S_x$ bounds the integrand by
$4(\Omega x^2)^{-1}\ln(1+x\Omega/(2\nu))$. Integration by parts bounds its
integral by $2\nu^{-1}[1+\ln(1+2\nu/\Omega)]$. The region $A_x\ge\Omega/2$ is
empty. Thus $\EE I\le8\nu^{-1}[1+\ln(\nu/\Omega)]$. Dyadic pairing and
monotonicity give
\begin{equation}
 \mathcal Q_\nu(R)\le
 2\left[\frac1{\nu+\Omega_R}+\frac{I}{\ln2}+\frac1{\nu+M_R}\right].
\end{equation}
Taking expectations yields a prefix with
$\mathcal Q_\nu(R)\le40\nu^{-1}[1+\ln(\nu/\Omega)]$. Apply this result with
$\widetilde a_m=a_m\one_{\{m\ge r_*\}}$, retaining the original distance indices.
Since $\xi>0$, $R(\xi)\ge r_*$. Adding back the earlier rates meets the original
moment target and only increases all denominators, which proves
Eq.~\eqref{eq:sm-prescribed}.

For the coarse rates set $d=\max(3K,Kr_*-K/2-1)$. Then exactly
$\Omega_*=(\beta K/4)F_d$. All microscopic distances in $(3K,d]$ belong to bins
below $r_*$. The rounding bound $j\le2Km$ gives $S_d\le S_{3K}+8Z/\beta$.
Equation~\eqref{eq:sm-tailbound} proves Eq.~\eqref{eq:sm-microprescribed}.

\subsection{A scale controlling moment and tail at once}
\label{sec:sm-simultaneous}

Assume $\sum_{j\ge0}1/U(2^j)=\infty$. For an integer $R_0\ge3$ with
$M(R_0)\ge\nu$ and $0<\epsilon\le1$, we show that there is $x\ge R_0$,
$R=\lceil x\rceil$, such that
\begin{align}
 U(32x^2)&\le4U(x),&
 g(x)&\le U(x)/2,&
 g(11x^2)&\le\epsilon U(x),\label{eq:sm-simultaneous}\\
 \mathcal Q_\nu(R)&\le(500+1000/\epsilon)\mathcal Q_\nu(R_0).
 \label{eq:sm-selectorbudget}
\end{align}
For the microscopic rates and $B=4R^2K$, these imply
\begin{equation}
 \beta(S_{3B}-S_{3K})\le64M(R),\qquad
 \beta S_{3B}\ge2M(R),\qquad
 3BF_{3B}\le10\epsilon S_{3B}.
 \label{eq:sm-microselection}
\end{equation}

Set $t_0=\ln R_0$, $v(t)=U(e^t)$, $v_0=v(t_0)$, and
$I(t_1)=\int_{t_0}^{t_1}dt/v(t)$. Then $v'(t)=g(e^t)$ almost everywhere and
$I(t_1)\to\infty$ as $t_1\to\infty$. Let $D_0=\{t:v(2t+\ln32)\le4v(t)\}$. If there
is no point satisfying Eq.~\eqref{eq:sm-simultaneous} before $t_1$, almost every
point of $D_0\cap[t_0,t_1]$ lies in $D_1=\{v'>v/2\}$ or
$D_2=\{v'(2t+\ln11)>\epsilon v(t)\}$. The $dt/v(t)$ measure of $D_1$ is at most
$2/v_0$, by integrating $2v'/v^2$. On $D_0$ one has $v(2t+\ln11)\le4v(t)$;
changing variables in the same derivative integral bounds the measure of
$D_0\cap D_2$ by $8/(\epsilon v_0)$. Consequently
\begin{equation}
 \int_{D_0\cap[t_0,t_1]}\frac{dt}{v(t)}
 \le(2+8/\epsilon)/v_0 .
\end{equation}
On the complement of $D_0$, $1/v(2t+\ln32)<1/(4v(t))$. A change of variables also
bounds the integral of $1/v(2t+\ln32)$ from below by
$[I(t_1)-I(2t_0+\ln32)]/2$. Hence the absence of a good point forces
\begin{equation}
 I(t_1)\le I_*=
 [2t_0+2\ln32+8+32/\epsilon]/v_0 .
\end{equation}
Choose a finite $t_1$ with $I(t_1)=I_*+1/v_0$. A good point exists before $t_1$,
outside the derivative null sets. Rounding $x=e^t$ up to $R$ adds at most $1/v_0$
to the full-profile integral.

At this point $M(x)\ge U(x)/2$. Since $R\le3x/2$, $RG(R)\le3M(R)/2$. For
$1\le y\le R$, $U_R(y)\ge(y/R)M(R)$, whence $U(y)\le(5/2)U_R(y)$. On the initial
interval, a dyadic bound gives
\begin{equation}
 \int_1^{R_0}\frac{dy}{y[\nu+U(y)]}\le\mathcal Q_\nu(R_0),\qquad
 \mathcal Q_\nu(R_0)\ge\frac{2\log_2R_0+1}{\nu+M(R_0)} .
\end{equation}
Because $M(R_0)\ge\nu$, these inequalities imply
$2t_0/v_0\le2\mathcal Q_\nu(R_0)$ and $1/v_0\le2\mathcal Q_\nu(R_0)$. Combining
the integral costs above gives
\begin{equation}
 \int_1^R\frac{dy}{y[\nu+U(y)]}
 \le(40+64/\epsilon)\mathcal Q_\nu(R_0).
\end{equation}
Dyadic pairing now yields
\begin{align}
 \mathcal Q_\nu(R)&\le4/\nu+4\int_1^R\frac{dy}{y[\nu+U_R(y)]}\nonumber\\
 &\le4/\nu+10\int_1^R\frac{dy}{y[\nu+U(y)]}
 \le(408+640/\epsilon)\mathcal Q_\nu(R_0),
\end{align}
where the $j=0$ term gives $\mathcal Q_\nu(R_0)\ge1/(2\nu)$. This proves
Eq.~\eqref{eq:sm-selectorbudget}.

Finally $12R^2+1\le32x^2$, so $M(12R^2+1)\le8M(R)$. For microscopic distances
up to $3B$, the rounding inequality $j\le2Km$ gives
$\beta(S_{3B}-S_{3K})\le8M(12R^2)\le64M(R)$. All retained distances lie below
$3B$, and $Km\le2j$ gives $\beta S_{3B}\ge2M(R)$. Every distance beyond $3B$ has
rounded index at least $12R^2>11x^2$. Thus
\begin{equation}
 \beta\,3BF_{3B}
 \le48R^2G(11x^2)
 \le(108/11)g(11x^2)
 <10\epsilon U(x)\le20\epsilon M(R).
\end{equation}
Combining with the preceding lower bound proves the last assertion of
Eq.~\eqref{eq:sm-microselection}.

\section{Fixed-temperature renewal and infinite volume}
\label{sec:sm-renewal}

We now close the sufficient direction. Reduce first to support of greatest
common divisor one and assume the reciprocal series diverges, $\mathcal R=\infty$,
since Sec.~\ref{sec:sm-summable} covers the other branch. For each fixed $K$,
the full coarse profile of the rates~\eqref{eq:sm-coarserates} satisfies
\begin{equation}
 U(n)=(\beta/4)E_{Kn}+O_K(1).
 \label{eq:sm-coarseasymptotic}
\end{equation}
Indeed, removing lengths up to $3K$ has a bounded moment cost, and rounding $j/K$
changes $\min(j,Kn)$ by at most $K/2$ per unit coupling. Summability bounds the
total error independently of $n$. Thus the coarse reciprocal series also
diverges; fixed rescaling only shifts dyadic comparisons. Its first moment is
infinite.

\subsection{One renewal step restores every budget}
\label{sec:sm-onestep}

Fix once and for all
\begin{equation}
 C_0=10^6,\qquad \beta\ge10^{40}(C+1),\qquad
 \epsilon=(160C_0)^{-1}.
\end{equation}
Here and below numerical couplings are expressed in the fixed energy unit of
Sec.~\ref{sec:sm-uniqueness}. At an admissible even $K$, put $S=S_{3K}$ and
$D=3KF_{3K}$. Suppose the weak and strong internal failure bounds are
\begin{equation}
 u_w\le e^{-\beta S/C_0},\qquad
 u_s\le e^{-\tau\beta S/C_0},
 \qquad \theta\ge24/25,\quad0<\tau\le1 .
\end{equation}
For $0<\delta\le1/32$, impose
\begin{align}
 \beta S&\ge2C_0\ln12,&
 \beta S&\ge524288C_0\ln(1/\delta),\nonumber\\
 \tau\beta S&\ge2C_0\ln(e/\delta),&
 \beta S&\ge C_0\ln1027,\nonumber\\
 \delta\beta S&\ge C_0\ln(3+\delta^{-2}),\label{eq:sm-renewalH}\\
 1+\ln[\beta S/(16C_0)]&\le CS,&
 4\ln(3+1/\tau)+2&\le CS .
 \label{eq:sm-renewalL}
\end{align}
Set $b=\beta S/(4C_0)$ and $\nu=b/4$. The full coarse mass is
$A=\beta D/12\le\beta S/(192C_0)$, so $b\ge16A$ and $\nu\ge A$. The weak-seed
condition of Sec.~\ref{sec:sm-transfer} follows from Eq.~\eqref{eq:sm-renewalH}.

Apply the prescribed-moment estimate of Sec.~\ref{sec:sm-prescribed} with target
$Z=\beta S$. For sufficiently large $K$, $\beta K\ge4$.
Equations~\eqref{eq:sm-microprescribed} and \eqref{eq:sm-renewalL} give
$\mathcal Q_\nu(R_a)\le400CS/\nu$. Pad the prefix to
$R_0=\max(R_a,3,\lceil1/\tau\rceil)$. Adding rates only increases existing
denominators; there are at most $2\log_2\max(3,\lceil1/\tau\rceil)+2$ new dyadic
terms, each at most $1/\nu$. Therefore $\mathcal Q_\nu(R_0)\le401CS/\nu$, while
$M(R_0)\ge\beta S$.

Apply the selection of Sec.~\ref{sec:sm-simultaneous} and put
$C_{\rm sel}=500+1000/\epsilon$. The profile budget becomes
\begin{equation}
 \frac{16}{b}+8\mathcal Q_\nu(R)
 \le(4+3208C_{\rm sel})\,16C_0C/\beta<10^{-4}.
\end{equation}
For the strong-average ceiling use $M_R\le AR$, $y_s\ge\tau\beta S/C_0$, and
$R\ge1/\tau$. Since $N=4R^2$,
$A/(512Ry_s)\le1/(98304R\tau)\le1/98304$; in particular $M_R\le128Ny_s$. All
other conditions in Eq.~\eqref{eq:sm-transferbudget} follow from
Eq.~\eqref{eq:sm-renewalH}.

For the selected parent $B=4R^2K$, write $S'=S_{3B}$ and $D'=3BF_{3B}$.
Equation~\eqref{eq:sm-microselection} and $M_R\ge\beta S$ give
\begin{equation}
 \beta S'\le65M_R,\qquad S'\ge2S,\qquad
 D'\le S'/(16C_0).
 \label{eq:sm-renewedmoment}
\end{equation}
The inequalities $1/8192-65/C_0\ge1/C_0$ and $1/256-65/C_0\ge1/C_0$, together
with Eq.~\eqref{eq:sm-renewalH}, absorb the two prefactors of the transfer
estimate. The renewed estimates are therefore
\begin{equation}
 u'_w\le e^{-\beta S'/C_0},\qquad
 u'_s\le e^{-\delta\beta S'/C_0},\qquad
 \theta'=\theta(1-2\delta),\quad\tau'=\delta .
\end{equation}
Both the admissible tail ratio~\eqref{eq:sm-admissible} and the weak estimate are
restored.

\subsection{A single finite initialization serves every scale}
\label{sec:sm-initialization}

Choose $\delta_k=[1000(k+1)^2]^{-1}$, $\tau_0=1$, and $\tau_k=\delta_{k-1}$ for
$k\ge1$. Starting at $\theta_0=0.99$, the density product remains above
$0.99(1-0.004)>0.98>24/25$. The lower bounds on $S$ in
Eq.~\eqref{eq:sm-renewalH} and the second part of Eq.~\eqref{eq:sm-renewalL}
grow at most polynomially times a logarithm in $k$. If their maximum with one is
$L_k$, then $\sup_k2^{-k}L_k<\infty$. The first part of
Eq.~\eqref{eq:sm-renewalL} holds for all sufficiently large $S$, since
$CS-\ln S\to\infty$ and is eventually increasing. Set
$L_*=\sup_{k\ge0}2^{-k}L_k<\infty$. Choose an admissible $K_0$ with
$S_{3K_0}\ge L_*$, beyond the eventual threshold for the first logarithmic
inequality, and with $3K_0\ge R_*$. Then $S_{3K_k}\ge2^kS_{3K_0}\ge L_k$ meets
every future requirement. The starting scale may also be chosen beyond the finite
connectivity threshold specified next.

To start both cluster estimates, choose a finite subset $\mathcal S$ of the
positive support with greatest common divisor one, and write
$f_-=\min\mathcal S$, $f_+=\max\mathcal S$, $j_*=\min_{f\in\mathcal S}J(f)>0$.
Every interval of $K_0\ge f_-+f_+$ vertices is connected by steps from
$\mathcal S$: length-$f_-$ edges connect each residue chain, and edges from $r$ to
$r+f$, $0\le r<f_-$, connect the residue graph. Choose one finite $\gamma\ge\beta$
such that
\begin{equation}
 (K_0-1)e^{-\gamma j_*}\le e^{-\beta S_{3K_0}/C_0}.
\end{equation}
Use the fixed auxiliary independent-bond law with rates $\gamma J(r)$ for
$r<K_0$ and $\beta J(r)$ for $r\ge K_0$. A spanning tree proves both base
estimates. Every subsequent fresh distance exceeds $3K\ge3K_0$, so its rate is
exactly $\beta J(r)$. All internal events at every stage belong to this single
law; no microscopic rate is used as two independent copies. Auxiliary comparison
couplings at successive stages need only establish these marginal estimates.

\subsection{Passage to order}
\label{sec:sm-passage}

The induction gives $K_k\to\infty$, $\theta_k>0.98$, and
$u_{s,k}\le\exp[-\tau_k\beta S_{3K_k}/C_0]\to0$. Average over a uniform site in an
interval of length $K_k$. Translation invariance of the full auxiliary law
yields
\begin{equation}
 \PP(|\mathcal C(0)|\ge\theta_kK_k)
 \ge\theta_k(1-u_{s,k}).
\end{equation}
The threshold diverges, so $\PP(|\mathcal C(0)|=\infty)\ge\lim_k\theta_k>0.98$.
No nested-giant coupling or independence between scales is needed.

The independent law with rates $\gamma J$ dominates this auxiliary law. In the
wired random-cluster representation $\phi^w_\gamma$ of the Ising chain at inverse
temperature $\gamma$, with edge probabilities $p_\gamma(r)=1-e^{-2\gamma J(r)}$,
each edge's minimum conditional opening probability is
$\tanh(\gamma J(r))\ge1-e^{-\gamma J(r)}$. Finite-volume monotone coupling
followed by the wired limit and the Edwards--Sokal
identity~\cite{Edwards_1988go} gives $m_+(\gamma)\ge0.98$. Ferromagnetic
monotonicity gives order at every larger inverse temperature. The residue-class
reduction of Sec.~\ref{sec:sm-cut} extends the result to arbitrary nonzero
support greatest common divisor, and the zero coupling is disordered. Together
with Sec.~\ref{sec:sm-summable}, this proves that $C_->0$ implies
$m_+(\beta)>0$ at some finite $\beta$.

The construction establishes existence at a finite coupling multiplier; it does
not provide a sharp numerical threshold or determine critical exponents.
Isoperimetry also enters critical-exponent bounds for long-range percolation
with a power-law upper envelope~\cite{Baumler_2024il}, and power-law kernels
below the upper-critical dimension have no infinite cluster at
criticality~\cite{Berger_2002tr,Hutchcroft_2021pl}; those results concern
critical behavior rather than a criterion for the existence of an ordered phase.
The double-exponential growth of the susceptibility of the marginal hierarchical
model as $\beta\to\infty$~\cite{Easo_2024de} is likewise a quantitative
statement about one kernel.
For regular power-law bond probabilities, monotone-path second-moment bounds
give explicit estimates of the percolation threshold~\cite{Amit_2026eu}, which
numerical studies locate~\cite{Gori_2017od}; the bounds'
renewal-overlap condition differs from the logarithmic condition $C_->0$ used
here. The construction above addresses arbitrary summable tails, including the
marginal regime $\mathcal R=\infty$.

\section{Random-cluster and Potts chains}
\label{sec:sm-potts}

The wired random-cluster measure $\phi^w_{t,q}$ has edge parameters
$p_{ij}=1-e^{-tJ(|i-j|)}$ and fixed finite $q\ge1$. Conditionally on the other
edges, an edge of rate $x=tJ(r)$ opens with probability between
$(1-e^{-x})/[1+(q-1)e^{-x}]$ and $1-e^{-x}$. For $q=1$ these coincide. For
$q>1$, the lower value is at least $1-e^{-x/q}$: setting $y=e^{-x/q}$, the
difference has the sign of $1-qy^{q-1}+(q-1)y^q\ge0$ for $0\le y\le1$.
Finite-volume monotone coupling and wired limits therefore give
\begin{equation}
 \PP_{t/q}\preceq\phi^w_{t,q}\preceq\PP_t,\qquad
 t_c(1)\le t_c(q)\le q\,t_c(1).
 \label{eq:sm-pottscomparison}
\end{equation}
Here $\PP_t$ denotes independent bonds of rates $tJ$, and $t_c(q)$ is the
threshold for positive infinite-cluster probability. Path counting excludes
independent percolation when $2t\Sigma_J<1$. The Ising construction of
Secs.~\ref{sec:sm-summable}--\ref{sec:sm-renewal} proves $t_c(1)<\infty$ when
$C_->0$. If independent bonds at $t$ percolate, the wired Ising law at inverse
temperature $t$ percolates by the comparison in Sec.~\ref{sec:sm-passage}.
The uniqueness criterion of Sec.~\ref{sec:sm-uniqueness} therefore excludes
independent percolation at any finite $t$ when $C_-=0$. Consequently,
\begin{equation}
 0<t_c(q)<\infty\quad\Longleftrightarrow\quad C_->0.
 \label{eq:sm-pottscriterion}
\end{equation}
For integer $q\ge2$, Edwards--Sokal coloring~\cite{Edwards_1988go} gives
\begin{equation}
 \frac{q\langle\one_{\{s_0=1\}}\rangle_1-1}{q-1}
 =\phi^w_{\beta,q}(0\leftrightarrow\infty)
\end{equation}
for the Potts Hamiltonian $-\sum_{i<j}J(j-i)\delta_{s_i,s_j}$ and boundary color
1. Its intensity is $t=\beta$, whereas the Ising Hamiltonian
$-\sum_{i<j}J(j-i)\sigma_i\sigma_j$ has $t=2\beta$. These are applications of the
standard random-cluster comparison~\cite{Grimmett_Book}, with no conclusion
uniform as $q\to\infty$ and no assertion about the continuity of the transition.

\section{Pressure, magnetic response and finite-range bounds}
\label{sec:sm-response}

Throughout this section the inverse temperature is $\beta$ and the
dimensionless field is $h=\beta H$. Write $J^{(R)}(r)=J(r)\one_{\{r\le R\}}$, so
that $F_R=\sum_{r>R}J(r)$ is the removed weight. The pressure $p(h)$ is the
thermodynamic limit of $|\Lambda|^{-1}\ln Z_\Lambda$; no factor $1/\beta$ is
included. We write $p_R$ and $m_R(h)$ for the pressure and the magnetization of
the truncated coupling $J^{(R)}$. We use free-boundary intervals for the
finite-volume arguments. The finite-range zero-field Gibbs state is unique and
has finite susceptibility
$\chi_R(\beta)=\sum_x\langle\sigma_0\sigma_x\rangle_{J^{(R)},\beta}$. These facts
follow from the primitive finite-state transfer matrix, including when some
retained couplings vanish. For $R=0$ the spins are independent and $\chi_0=1$.

\subsection{Truncation changes the pressure uniformly and the field recovers the magnetization}
\label{sec:sm-pressure}

The weight of all removed pairs in an interval of $n$ sites is at most $nF_R$.
The energy bound and Griffiths monotonicity give
\begin{equation}
 0\le p(h)-p_R(h)\le\beta F_R\qquad(h\in\mathbb R).
 \label{eq:sm-pressure}
\end{equation}
Monotonicity applies at nonnegative field, and spin-flip symmetry covers negative
field. Boundary effects vanish per site because $E_n/n\to0$ by summability. If
$m_*=m_+(\beta)>0$, the standard pressure identity gives $p'_+(0)=m_*$. Indeed,
the positive-field derivative tends to the right derivative by convexity; the
plus states decrease to the maximal zero-field Gibbs state by monotonicity,
compactness, and the Gibbs specification. Consequently,
$p_R(h)-p_R(0)\ge m_*h-\beta F_R$ for $h>0$. Convexity of $p_R$ yields
\begin{equation}
 m_R(h)\ge m_*-\frac{\beta F_R}{h}.
 \label{eq:sm-field}
\end{equation}
Monotonicity in $R$ and comparison with $m_+(\beta,h)$ then imply
\begin{equation}
 \lim_{h\downarrow0}\lim_{R\to\infty}m_R(h)=m_*,\qquad
 \lim_{R\to\infty}\lim_{h\downarrow0}m_R(h)=0.
\end{equation}
Every magnetization here is already in the thermodynamic limit at fixed $R,h$; no
joint sample-size and cutoff limit is asserted.

\subsection{Restoring the tail raises the susceptibility by a controlled factor}
\label{sec:sm-interpolation}

Fix $R<R'\le\infty$ and set $\Delta J=J^{(R')}-J^{(R)}$, with $J^{(\infty)}=J$
and $F_\infty=0$. In a finite free-boundary interval $\Lambda$, interpolate
$J_s=J^{(R)}+s\,\Delta J$, $0\le s\le1$. At zero field put
\begin{equation}
 \Gamma_{ij}(s)=\langle\sigma_i\sigma_j\rangle_s,\qquad
 X_\Lambda(s)=\max_{i\in\Lambda}\sum_{j\in\Lambda}\Gamma_{ij}(s).
\end{equation}
The Lebowitz inequality~\cite{Lebowitz_1974ga} states
\begin{equation}
 \langle\sigma_i\sigma_j\sigma_u\sigma_v\rangle_s
 -\Gamma_{ij}(s)\Gamma_{uv}(s)
 \le\Gamma_{iu}(s)\Gamma_{jv}(s)+\Gamma_{iv}(s)\Gamma_{ju}(s).
\end{equation}
It holds also for coincident indices. Differentiating with respect to $s$ and
using the pair convention $u<v$ therefore gives
\begin{equation}
 \partial_s\Gamma_{ij}(s)\le\beta\sum_{u<v}\Delta J(|u-v|)
 [\Gamma_{iu}(s)\Gamma_{jv}(s)+\Gamma_{iv}(s)\Gamma_{ju}(s)].
\end{equation}
The sum of added couplings incident to any site is at most $2(F_R-F_{R'})$.
Summing over $j$ and taking the upper right Dini derivative of the finite maximum
yields
\begin{equation}
 D^+X_\Lambda(s)\le 2\beta(F_R-F_{R'})X_\Lambda(s)^2.
 \label{eq:sm-dini}
\end{equation}
Adding sites and bonds increases correlations, so $X_\Lambda(0)\le\chi_R(\beta)$.
Integration of Eq.~\eqref{eq:sm-dini} gives, whenever
$2\beta(F_R-F_{R'})\chi_R<1$,
\begin{equation}
 X_\Lambda(1)\le
 \frac{\chi_R}{1-2\beta(F_R-F_{R'})\chi_R}.
 \label{eq:sm-finitecomparison}
\end{equation}
The maximum row sum is essential: free boundaries do not give identical row sums
in a finite interval.

For finite $R'$, taking increasing intervals and then summing correlations
proves
\begin{equation}
 \chi_R\le\chi_{R'}\le
 \frac{\chi_R}{1-2\beta(F_R-F_{R'})\chi_R}.
 \label{eq:sm-chicomparison}
\end{equation}
For $R'=\infty$, the Griffiths--Hurst--Sherman
inequality~\cite{Griffiths_1970co,Lebowitz_1974ga} bounds the finite-volume
pressure increment at $h\ge0$ by $X_\Lambda(1)h^2/2$. Thus $2\beta F_R\chi_R<1$
gives a volume-independent quadratic bound on $p(h)-p(0)$ and forces
$m_+(\beta)=p'_+(0)=0$. An ordered temperature must therefore satisfy
\begin{equation}
 \chi_R(\beta)\ge\frac{1}{2\beta F_R}.
 \label{eq:sm-orderedchi}
\end{equation}
The argument proves the non-strict endpoint without an assumption about
continuity at an interpolation critical point. For an ordered coupling,
$F_R>0$ at every finite $R$. For the dyadic chain $J_a(2^k)=a\ln2\,2^{-k}$ one
has $F_R=a\ln2\,2^{-\lfloor\log_2R\rfloor}\le2a\ln2/R$, hence
$\chi_R\ge R/(4\beta a\ln2)$ at every ordered temperature.

\subsection{The cutoff criterion is complete and monotone}
\label{sec:sm-cutoff}

Define $f_R(\beta)=2\beta F_R\chi_R(\beta)$. For fixed $R$, $\chi_R$ is analytic
at finite $\beta$, nondecreasing, and at least one. If $F_R>0$, $f_R$ is
continuous and strictly increasing from zero to infinity. Thus it has a unique
positive root $f_R(\beta_R)=1$. If $F_R=0$, define $\beta_R=\infty$.

If $f_R(\beta)<1$, continuity gives $f_R(\beta')<1$ for some $\beta'>\beta$. The
preceding argument excludes order at $\beta'$, so monotonicity of the
magnetization gives $\beta<\beta_c$. Conversely, sharpness for transitive Ising
models with arbitrary nonnegative invariant
couplings~\cite{Aizenman_1987tp,Duminil-Copin_2015an,Duminil-Copin_2018ct} gives
finite susceptibility $\chi(\beta)$ for $\beta<\beta_c$. Theorem~2.1(2) of
Ref.~\cite{Duminil-Copin_2015an} applies here by taking the underlying locally
finite transitive graph to be the nearest-neighbor line; its interaction is
allowed between arbitrary pairs of vertices. Since $\chi_R\le\chi$ and $F_R\to0$,
every such $\beta$ satisfies $f_R(\beta)<1$ for all sufficiently large $R$.
Therefore
\begin{equation}
 \beta<\beta_c\quad\Longleftrightarrow\quad
 \exists R:\ f_R(\beta)<1.
 \label{eq:sm-complete}
\end{equation}

For monotonicity, let $R<R'$ with $F_{R'}>0$. At $\beta=\beta_R$,
\begin{equation}
 1-2\beta_R(F_R-F_{R'})\chi_R(\beta_R)=F_{R'}/F_R>0.
\end{equation}
Equation~\eqref{eq:sm-chicomparison} gives $f_{R'}(\beta_R)\le1$ and hence
$\beta_{R'}\ge\beta_R$. If $F_{R'}=0$ the same conclusion follows from the
definition. Equation~\eqref{eq:sm-complete} then yields
\begin{equation}
 \beta_R\uparrow\beta_c,
 \label{eq:sm-rootlimit}
\end{equation}
where infinity is allowed on both sides. No monotonicity of $J(r)$ is used and no
rate of convergence is implied. This is a consequence of classical correlation
inequalities and sharpness, in the tradition of finite-volume
criteria~\cite{Simon_1980ci,Lieb_1980ar,Duminil-Copin_2015an}.

There is also an error bound in the certified subcritical regime. When
$f_R(\beta)<1$, uniqueness follows as above. Taking the infinite-volume limit in
Eq.~\eqref{eq:sm-finitecomparison} with $R'=\infty$ gives
\begin{equation}
 \chi_R\le\chi\le\frac{\chi_R}{1-f_R},\qquad
 0\le\chi-\chi_R\le\frac{f_R}{1-f_R}\chi_R.
 \label{eq:sm-error}
\end{equation}
Thus a small $f_R$ controls the susceptibility truncation error, while in the
ordered regime $f_R\ge1$ for every cutoff. Certified numerical evaluation of
$\chi_R$ and the choice of an efficient cutoff are separate computational issues;
no such computation is used in the main text. Finite-range scaling studies
extrapolate truncated chains numerically~\cite{Glumac_1989fr,Andrade_2005ts,Aoki_2008af};
the roots $\beta_R$ give rigorous lower bounds that increase to $\beta_c$.

\subsection{The nearest-neighbor cutoff gives a Lambert-function bound}
\label{sec:sm-lambert}

The first nontrivial bound requires only the exact nearest-neighbor
susceptibility, $\chi_1(\beta)=e^{2\beta J(1)}$. For $J(1)>0$ and $F_1>0$,
setting $x=2\beta J(1)$ turns $f_1(\beta)=1$ into $xe^x=J(1)/F_1$, so
\begin{equation}
 \beta_1=\frac{W[J(1)/F_1]}{2J(1)}\le\beta_c,
 \label{eq:sm-firstcut}
\end{equation}
where $W(x)e^{W(x)}=x$ is the positive Lambert function. For $J(1)=0$ and
$F_1>0$, the bound is $\beta_1=1/(2F_1)$. The independent-spin starting point
$R=0$ gives $\beta_0=1/(2\Sigma_J)$, the high-temperature bound of
Sec.~\ref{sec:sm-uniqueness}; including correlations in the retained chain and
then extending its range improves this initial bound. For the dyadic chain
$J_a(2^k)=a\ln2\,2^{-k}$, $J_a(1)=F_1=a\ln2$ and
\begin{equation}
 \beta_c\ge\frac{W(1)}{2a\ln2}.
 \label{eq:sm-dyadiclambert}
\end{equation}
This explicit finite-temperature estimate coexists with the criterion
guaranteeing $\beta_c<\infty$. Higher cutoffs require larger transfer matrices,
so the convergence~\eqref{eq:sm-rootlimit} is a rigorous characterization rather
than a claim of uniform computational efficiency.

\subsection{A finite-range observation cannot certify order}
\label{sec:sm-observation}

Fix nonnegative couplings $\mathcal K(1),\ldots,\mathcal K(R)$ with $R<\infty$,
and set
\begin{equation}
 \mathcal K^0(r)=\mathcal K(r)\one_{\{r\le R\}},\qquad
 \mathcal K^a(r)=\mathcal K^0(r)+a r^{-2}\one_{\{r>R\}},\qquad a>0.
\end{equation}
Writing $Z_R=\sum_{r>R}r^{-2}$, the two couplings agree at every observed
distance and satisfy $\|\mathcal K^a-\mathcal K^0\|_1=aZ_R$. The finite-range
chain $\mathcal K^0$ has $C_-=0$. For $N>R$,
\begin{equation}
 E_N(\mathcal K^a)=\sum_{r\le R}r\mathcal K(r)
       +a\sum_{R<r\le N}\frac1r
       +aN\sum_{r>N}\frac1{r^2}
       =a\ln N+O_{\mathcal K,R,a}(1).
\end{equation}
Hence $C_-(\mathcal K^a)=a>0$, and the ordering criterion gives
$0<\beta_c(\mathcal K^a)<\infty$, whereas $\mathcal K^0$ is disordered at every
finite $\beta$. Taking $aZ_R<\epsilon$ gives arbitrarily close couplings with
different ordering classifications. Their pressures obey
$0\le p_{\mathcal K^a}(h)-p_{\mathcal K^0}(h)\le\beta aZ_R$ for all fields. This
is a consequence of the criterion and the pressure estimate~\eqref{eq:sm-pressure},
not a separate transition mechanism. In one dimension every finite truncation is
disordered, whereas in dimensions $d\ge2$ long enough truncations of nonsummable
long-range percolation, with support not contained in a line, still
percolate~\cite{Friedli_2004ol,Friedli_2006ot,Baumler_2024to,Hartarsky_2026wi}.

In fact, order for this particular extension follows already from the
inverse-square theorem of Fr\"ohlich and Spencer~\cite{Frohlich_1982tp}. Set
$L=R+1$ and retain only bonds within the sublattice $L\ZZ$. Its distinct sites
have separations $Ln>R$, so the retained couplings are $a/(Ln)^2$. After
rescaling the lattice, this is an inverse-square chain with positive amplitude
$a/L^2$, which orders at sufficiently low temperature. Restoring the deleted
ferromagnetic bonds can only increase the plus-state magnetization. Thus neither
this example nor the density statement below requires the general sufficient
direction proved in Secs.~\ref{sec:sm-branches}--\ref{sec:sm-renewal}.

The construction also shows that the ordered and the disordered classes are both
dense in the cone of nonnegative summable couplings with its $\ell^1$ norm. Given
$J$ and $\epsilon>0$, first choose a truncation with $\|J-J^{(R)}\|_1<\epsilon/2$,
then add the above tail with $aZ_R<\epsilon/2$. Both approximants are within
$\epsilon$ of $J$. The truncation can be chosen beyond any prescribed set of
observed distances. In contrast, perturbations of finite weighted first moment
preserve $C_-$, because
$|E_N(J)-E_N(\widetilde J)|\le\sum_rr|J(r)-\widetilde J(r)|$.

For a fixed head $\mathcal K^0$, the ordering temperature of $\mathcal K^a$ tends
to zero as $a\downarrow0$. Indeed, at every fixed finite $\beta$, the head
susceptibility $\chi_{\mathcal K^0}(\beta)$ is independent of $a$ and finite.
Eventually $2\beta aZ_R\chi_{\mathcal K^0}(\beta)<1$, which implies
$\beta<\beta_c(\mathcal K^a)$ by Eq.~\eqref{eq:sm-complete}. Thus
$\beta_c(\mathcal K^a)\to\infty$. This approximation obstruction concerns the
existence of an ordered phase at some temperature; it does not assert order at a
fixed positive temperature under an arbitrarily weak tail.


\section{The magnetization gap}
\label{sec:smj-gap}

This section proves part~(i) of Theorem~2 of the main text, the magnetization gap. Let $J(r)\ge0$
be translation invariant with $\Sigma_J<\infty$ and $C_+<\infty$. Then for every
$0\le\beta<\infty$,
\begin{equation}
m_+(\beta)=0\qquad\text{or}\qquad 2\beta C_+\,m_+(\beta)^2\ge1 .
\label{eq:smj-gap}
\end{equation}
Since $m_+(\beta)<1$ at every finite $\beta$ [Eq.~\eqref{eq:smj-mless1}
below], an ordered $\beta$ satisfies $1\le2\beta C_+m_+(\beta)^2<2\beta C_+$,
so
\begin{equation}
m_+(\beta)>0\ \Longrightarrow\ \beta>\frac1{2C_+},
\label{eq:smj-nobelow}
\end{equation}
and $C_+=0$ excludes order at every finite temperature. Nothing is assumed
about $J$ beyond nonnegativity, summability and $C_+<\infty$: no
monotonicity, no decay exponent, no nearest-neighbor bond, no connected
support, and no bound on the coupling at individual scales. The ordering
criterion of the main text is not used; Eq.~\eqref{eq:smj-gap} proves no
existence of order.

\textit{Why this route.} The physical picture is that of Thouless, made
rigorous for inverse-square
couplings~\cite{Thouless_1969lr,Aizenman_1986do,Aizenman_1988do}: a long bond
ties two distant regions into the ordered cluster only if both ends belong
to it, which costs a factor of about $m_+^2$; if $m_+$ is small, the
effective long-range coupling is too weak to beat the logarithmic entropy
of large reversed regions. This picture was later turned into a
renormalization of ``unbridged'' blocks for the inverse-square
kernel~\cite{Duminil-Copin_2024lr}. That recursion is iterated along a
geometric sequence of scales and needs, at every scale $s$, bounded values
of the tail budget $sF_s$ and of the weight of bonds spanning a block. For a
general kernel these per-scale costs are not bounded even when $E_N$ grows
logarithmically: the pulse coupling of Sec.~\ref{sec:smj-pulse} has
$C_->0$, $C_+<\infty$ and $sF_s$ unbounded. The route below is open because
both costs, integrated over $\ln s$, reproduce the cumulative domain energy
exactly [Eqs.~\eqref{eq:smj-intB} and~\eqref{eq:smj-intM}]. We therefore
prove a recursion at every real scale $s$, and average its logarithm over
$\ln s$ instead of iterating it.

The proof runs in nine steps:
the random-cluster representation (Sec.~\ref{sec:smj-fk});
a zero-one law for the existence of an infinite cluster
(Sec.~\ref{sec:smj-zeroone}); the squared escape factor for two distant
boxes (Sec.~\ref{sec:smj-escape}); the blocks and their crossings
(Sec.~\ref{sec:smj-blocks}); the probability that a block is unbridged
(Sec.~\ref{sec:smj-unbridged}); the recursion at real scales
(Sec.~\ref{sec:smj-recursion}); a termination estimate
(Sec.~\ref{sec:smj-termination}); the logarithmic averages
(Sec.~\ref{sec:smj-average}); and the final inequality
(Sec.~\ref{sec:smj-final}).

\subsection{Random-cluster representation and escape probabilities}
\label{sec:smj-fk}

Let $E$ be the set of unordered pairs $e=\{x,y\}$ of distinct integers, with
length $|e|=|x-y|$ and parameter
$p_e=p_\beta(|e|)=1-e^{-2\beta J(|e|)}\in[0,1)$. A configuration
$\omega\in\{0,1\}^E$ declares $e$ open if $\omega_e=1$. For finite sets
$\Lambda\subset \bar\Lambda\subset\ZZ$, let $E_{\Lambda,\bar\Lambda}$ be the edges with
both endpoints in $\bar\Lambda$ and at least one in $\Lambda$, and let $\zeta$ be a
partition of $\bar\Lambda\setminus\Lambda$. The random-cluster measure with $q=2$ is
\begin{equation}
\phi^\zeta_{\Lambda,\bar\Lambda}(\omega)\propto
\prod_{e\in E_{\Lambda,\bar\Lambda}}p_e^{\omega_e}(1-p_e)^{1-\omega_e}\,
2^{k^\zeta(\omega)},
\label{eq:smj-fkmeasure}
\end{equation}
where $k^\zeta(\omega)$ is the number of open clusters on $\bar\Lambda$ after the
vertices in each block of $\zeta$ are identified. The wired law $w$ has one
block. Throughout, a path or a connection uses open edges only; the
identification by $\zeta$ enters the weight, never an event. Taking $\bar\Lambda\uparrow\ZZ$
gives the wired law $\phi^w_{\Lambda}$ on the edges touching $\Lambda$, with
the full infinite exterior, and taking $\Lambda\uparrow\ZZ$ gives the
infinite-volume wired measure $\phi^w=\phi^w_\beta$; both limits exist for
every summable $J$~\cite{Aizenman_1988do}.

We use four standard properties of the finite
laws~\cite{Fortuin_1972ot,Grimmett_Book}, valid for every $q\ge1$ and every
choice of $p_e\in[0,1)$, including $p_e=0$ at distances the coupling does
not support.
\begin{enumerate}
\item[(i)] \textit{Positive association (FKG).} Increasing events are
positively correlated, and so are decreasing events.
\item[(ii)] \textit{Single-edge bound.} Conditional on the states of all
other edges, $e$ is open with probability $p_e$ or
$p_e/[p_e+2(1-p_e)]$, hence at most $p_e$. Consequently any finite set of
edges is entirely closed with probability at least
$\prod(1-p_e)=\exp[-2\beta\sum J(|e|)]$, and by a decreasing limit the same
holds for an infinite set of edges with summable couplings.
\item[(iii)] \textit{Domain Markov property.} Fix a finite $D\subset\Lambda$.
Conditional on the states of all edges not touching $D$, the edges touching
$D$ are distributed as a random-cluster measure on those edges with a
boundary partition induced by the conditioning, possibly with some edges
absent. We call any such law a \emph{boundary law at $D$}. The same holds
when the conditioning set is revealed adaptively, edge by edge, since at
each stage the event that a given set has been revealed with given states is
determined by the states of that set.
\item[(iv)] \textit{Boundary maximum.} If $A$ is an increasing event
depending only on edges touching $D$, every boundary law $\phi'$ at $D$
satisfies $\phi'(A)\le\phi^w_D(A)$. Indeed, wiring the boundary increases
$A$; an absent edge is an edge conditioned closed, and restoring it
increases $A$ by FKG; enlarging $\bar\Lambda$ adds edges from $D$ to wired vertices,
which again increases $A$.
\end{enumerate}

The Edwards--Sokal coupling~\cite{Fortuin_1972ot,Edwards_1988go} identifies
the Ising model on $\Lambda$ with all spins outside $\Lambda$ fixed to $+1$
(all couplings kept) with $\phi^w_\Lambda$:
$\langle\sigma_x\rangle^+_{\Lambda,\beta}=\phi^w_\Lambda(x\leftrightarrow\Lambda^c)$.
For the boxes $\Lambda_R=[-R,R)\cap\ZZ$ define the escape probability
\begin{equation}
\rho_R(\beta)=\phi^w_{\Lambda_R}(0\leftrightarrow\Lambda_R^c)
=\langle\sigma_0\rangle^+_{\Lambda_R,\beta}.
\label{eq:smj-rho}
\end{equation}
By the Griffiths inequalities~\cite{Griffiths_1967ci,Kelly_1968gg}
$\rho_R$ decreases in $R$, and its limit is the plus-boundary
magnetization, $\rho_R(\beta)\downarrow m_+(\beta)$; moreover
$m_+(\beta)=\phi^w(0\leftrightarrow\infty)$~\cite{Aizenman_1988do}. The
magnetization is below one at finite $\beta$: by (ii), all edges at the
origin, of total coupling $2\Sigma_J$, are closed with probability at least
$e^{-4\beta \Sigma_J}$, and then the origin is isolated, so
\begin{equation}
m_+(\beta)\le1-e^{-4\beta \Sigma_J}<1 .
\label{eq:smj-mless1}
\end{equation}

\subsection{Local finiteness and the zero-one law}
\label{sec:smj-zeroone}

The proof ends by showing that a box cannot connect to distant sites with
probability close to one. This contradicts order only if order forces an
infinite cluster almost surely, not merely with positive probability. For a
general kernel the support graph has infinite degree and need not be
connected, so we prove the needed zero-one law directly.

\textit{Conditional domination.} For finite $D\subset\ZZ$ let
$\mathcal F_{D^c}$ be the $\sigma$-algebra of the edges with both endpoints
outside $D$. Every increasing event $f$ depending on finitely many edges
touching $D$ satisfies
\begin{equation}
\phi^w\bigl[f\,\big|\,\mathcal F_{D^c}\bigr]\le\phi^w_D(f)
\quad\text{almost surely.}
\label{eq:smj-conddom}
\end{equation}
To see this, fix a finite set $F$ of edges outside $D$ and states
$\eta\in\{0,1\}^F$ of positive probability. In a finite-volume law
$\phi^w_{\Lambda',\bar\Lambda}$ with $D\subset\Lambda'$, condition on all edges not
touching $D$; by (iii) the edges touching $D$ follow a boundary law at $D$
(edges from $D$ to vertices outside $\bar\Lambda$ are absent), so by (iv)
$\phi^w_{\Lambda',\bar\Lambda}(f\cap\{\omega_F=\eta\})\le\phi^w_D(f)\,
\phi^w_{\Lambda',\bar\Lambda}(\omega_F=\eta)$. Both sides are cylinder probabilities,
so the inequality survives $\bar\Lambda\uparrow\ZZ$ and then
$\Lambda'\uparrow\ZZ$, giving
$\phi^w(f\mid\omega_F=\eta)\le\phi^w_D(f)$. Letting $F$ increase to all
edges outside $D$, martingale convergence gives
Eq.~\eqref{eq:smj-conddom}. No infinite-volume Gibbs property is presupposed.

\textit{Tail triviality.} Let $\mathcal T=\bigcap_n\mathcal F_{\Lambda_n^c}$
be the $\sigma$-algebra of events that do not depend on any finite set of
vertices. For an increasing cylinder event $f$, reverse martingale
convergence and Eq.~\eqref{eq:smj-conddom} with $D=\Lambda_n$ give
$\phi^w[f\mid\mathcal T]\le\lim_n\phi^w_{\Lambda_n}(f)=\phi^w(f)$. Both sides
have expectation $\phi^w(f)$, so equality holds almost surely. The events
that a given finite set of edges is entirely open form a $\pi$-system
generating all events, so $\mathcal T$ is independent of every event,
including its own: each event of $\mathcal T$ has probability zero or one.

\textit{Local finiteness.} By (ii) every edge is open with probability at
most $p_e\le2\beta J(|e|)$ under $\phi^w$, so the expected open degree of a
vertex is at most $\sum_{y\ne x}2\beta J(|x-y|)=4\beta \Sigma_J<\infty$. Almost
surely every vertex has finite open degree.

\textit{Zero-one law.} Let $\mathcal I$ be the event that an infinite open
cluster exists, and $\mathcal I_n$ the event that the open edges with both
endpoints outside $\Lambda_n$ contain an infinite cluster. On a locally
finite graph, removing the finitely many vertices of $\Lambda_n$ cuts an
infinite cluster into finitely many pieces, at least one infinite; hence
$\mathcal I=\mathcal I_n$ almost surely for every $n$. The $\mathcal I_n$
decrease and $\bigcap_{n\ge k}\mathcal I_n$ is
$\mathcal F_{\Lambda_k^c}$-measurable for every $k$, so
$\bigcap_n\mathcal I_n\in\mathcal T$, and $\phi^w(\mathcal I)\in\{0,1\}$.
Therefore
\begin{equation}
m_+(\beta)>0\ \Longrightarrow\ \phi^w(\mathcal I)=1 .
\label{eq:smj-zeroone}
\end{equation}
Neither uniqueness of the infinite cluster nor connectedness of the support
of $J$ is used; a kernel supported on $d\ZZ$ is covered.

\subsection{The squared escape factor for two distant boxes}
\label{sec:smj-escape}

Let $\phi$ be any finite-volume law $\phi^\zeta_{\Lambda,\bar\Lambda}$, and let
$U=x+\Lambda_R$ and $V=y+\Lambda_R$ be two boxes inside $\Lambda$ with
$|x-y|>s>2R$; they are disjoint. Put $e=\{x,y\}$ and let $\mathcal E_x$ be the event
that $x$ is joined by an open path not using $e$ to a vertex outside $U$
(it depends only on edges touching $U$); define $\mathcal E_y$ likewise. Then
\begin{equation}
\phi(\mathcal E_x\cap \mathcal E_y)\le\rho_R(\beta)^2+8\beta R^2\,j_*(s-2R),
\qquad j_*(t)=\sup_{r>t}J(r).
\label{eq:smj-escape}
\end{equation}
Since $J$ is summable, $J(r)\to0$ and $j_*(t)\to0$ as $t\to\infty$.

The proof closes all bonds directly joining the two boxes and then exposes
one box at a time. Let $\mathcal D_{UV}$ be the event that every edge between $U$ and $V$
is closed, $e$ included. Under $\phi(\cdot\mid \mathcal D_{UV})$, which is the
random-cluster law with these edges absent, reveal all edges touching $U$;
$\mathcal E_x$ is determined by them. The edges touching $V$ are untouched by this
exposure, because the only edges touching both boxes are known to be closed.
Conditional on the exposure, and further on every edge not touching $V$,
the edges touching $V$ follow a boundary law at $V$, so by (iv) and
translation invariance $\phi(\mathcal E_y\mid \mathcal D_{UV},\text{edges touching }U)\le
\phi^w_V(\mathcal E_y)\le\rho_R$. The same argument without the first exposure gives
$\phi(\mathcal E_x\mid \mathcal D_{UV})\le\rho_R$. Hence $\phi(\mathcal E_x\cap \mathcal E_y\mid \mathcal D_{UV})\le\rho_R^2$. This
is sequential domination, not independence of the two boxes.

To return to the original law, in which $e$ and the other cross edges are
not forced closed,
$\phi(\mathcal E_x\cap \mathcal E_y)\le\phi(\mathcal E_x\cap \mathcal E_y\mid \mathcal D_{UV})+\phi(\mathcal D_{UV}^c)$, and by (ii)
$\phi(\mathcal D_{UV}^c)\le\sum_{u\in U,v\in V}p_\beta(|u-v|)$. There are $4R^2$ pairs,
each at distance at least $|x-y|-(2R-1)>s-2R$, and $p_\beta\le2\beta J$,
which gives Eq.~\eqref{eq:smj-escape}. The edge $e$ belongs to the
exceptional sum even though the events ignore it, because the original law
depends on it.

Call an open edge $e=\{x,y\}$ a \emph{bridge} if $\mathcal E_x\cap \mathcal E_y$ occurs. The
events $\mathcal E_x,\mathcal E_y$ do not depend on $e$, so conditioning on all other edges
and using (ii),
\begin{equation}
\phi(e\text{ is a bridge})\le p_e\,\phi(\mathcal E_x\cap \mathcal E_y)
\le p_e\bigl[\rho_R^2+8\beta R^2j_*(s-2R)\bigr].
\label{eq:smj-bridge}
\end{equation}
A long bond counts at full strength only when both its ends escape
locally, and that probability is close to $m_+^2$ once $R$ and $s$ are
large: this is the squared factor of the mechanism.

\textit{Parameters.} Fix $\beta>0$ and suppose $m=m_+(\beta)>0$; by
Eq.~\eqref{eq:smj-mless1}, $m<1$. Fix $\eta$ with $m<\eta<1$. Since
$\rho_R\downarrow m$, choose $R\ge1$ with $\rho_R(\beta)<\eta$, and then
$s_0>\max(2R,1)$ such that the bracket in Eq.~\eqref{eq:smj-bridge} is at
most $\eta^2$ for all $s\ge s_0$. For $s\ge s_0$ every edge $e$ whose two
boxes lie in $\Lambda$ and with $|e|>s$ is then a bridge with probability at most
$\eta^2p_e$. These choices do not involve the scale ratio $\lambda$
introduced below.

\subsection{Blocks, crossings and their geometry at real scales}
\label{sec:smj-blocks}

For real $s\ge1$ let $I_s=[-3s,3s)\cap\ZZ$, an interval of
$n_s=\lceil3s\rceil-\lceil-3s\rceil$ sites, $6s-1<n_s<6s+1$. An edge is
\emph{short} at scale $s$ if its length is at most $s$ and \emph{long}
otherwise. For an integer $z$, an \emph{$s$-crossing} of the block $z+I_s$
is an open path of short edges from a site $<z-3s$ to a site $\ge z+3s$
whose intermediate sites all lie in $z+I_s$. It depends only on the short
edges touching the block. Define
\begin{equation}
u(s)=\phi^w_{I_s}\bigl(I_s\text{ has no }s\text{-crossing}\bigr),
\qquad q(s)=-\ln u(s).
\label{eq:smj-u}
\end{equation}
By (iii) and (iv), under every boundary law at $z+I_s$, and under $\phi^w$
by Eq.~\eqref{eq:smj-conddom}, the block $z+I_s$ fails to be crossed with
probability at least $u(s)$. Closing all edges touching $I_s$, of total
coupling at most $2n_s\Sigma_J$, prevents a crossing, so by (ii)
\begin{equation}
0\le q(s)\le4\beta(6s+1)\Sigma_J .
\label{eq:smj-qbound}
\end{equation}
The block and the set of short lengths change only when $3s$ or $s$
crosses an integer, so $q$ is a step function with finitely many steps on
each bounded interval; it is measurable and locally bounded.

Two exact counts fix the costs. First, the edges of length $r$ that
\emph{span} a block, i.e.\ join a site $<z-3s$ to a site $\ge z+3s$, have
left endpoints filling an integer interval of length $(r-n_s)_+$. A spanning
edge is therefore long ($r>n_s>s$), and since $(r-n_s)_+\le r$, the total
coupling of spanning edges of length at most $\lambda s$ is at most
\begin{equation}
\mathcal M(s,\lambda)=\sum_{s<r\le\lambda s}rJ(r).
\label{eq:smj-M}
\end{equation}
Second, a set of $k$ sites is touched by long edges of total coupling at
most $2kF_s$, with $F_s=\sum_{r>s}J(r)$, since each site sees $F_s$ on each
side.

\subsection{The probability that a block is unbridged}
\label{sec:smj-unbridged}

Fix a real scale ratio $\lambda\ge36$ and a real $s\ge s_0$. The large
block is $I_{\lambda s}$, and we work in $\phi^w_{I_{\lambda s},\bar\Lambda}$ for
finite $\bar\Lambda$, sending $\bar\Lambda\uparrow\ZZ$ at the end. Put
$b_s=\lceil9s\rceil$, so $9s\le b_s\le10s$, and take as
\emph{candidates} the blocks $z+I_s$ with $z\in b_s\ZZ$ and
$|z|+3s\le\lambda s$. Their number is
\begin{equation}
\kappa=2\left\lfloor\frac{(\lambda-3)s}{b_s}\right\rfloor+1,
\qquad \frac{\lambda}{10}\le\kappa\le\lambda,
\label{eq:smj-nu}
\end{equation}
because $\kappa\ge2(\lambda-3)/10-1=\lambda/5-8/5\ge\lambda/10$ for
$\lambda\ge16$, and $\kappa\le2(\lambda-3)/9+1\le\lambda$. Distinct candidates
are more than $b_s-6s\ge3s$ apart, so no short edge touches two of them.

Each candidate has the \emph{guard} $G_z=z+([-5s,5s)\cap\ZZ)$, with at
most $10s+1\le11s$ sites. A candidate is \emph{unbridged} if (a) every long
edge touching $G_z$ is closed, and (b) no edge of length at most $\lambda s$
spanning $z+I_s$ is a bridge. Condition (a) excludes long edges of every
length, not only those up to $\lambda s$; this is what keeps the later
exposure away from the candidate.

The endpoints of a spanning edge of length at most $\lambda s$ lie in
$[-2\lambda s+6s,\,2\lambda s-6s]$, because $|z|\le\lambda s-3s$; their
boxes of radius $R<s/2$ lie inside $I_{\lambda s}$, so
Eq.~\eqref{eq:smj-bridge} applies to them with bound $\eta^2p_e$. The
guard lies inside $I_{\lambda s}$ as well. The complement of a bridge and the
closure of an edge are decreasing events, so FKG gives
\begin{equation}
\phi\bigl(z+I_s\text{ unbridged}\bigr)
\ge\prod_{\substack{e\text{ spans }z+I_s\\|e|\le\lambda s}}(1-\eta^2p_e)
\prod_{\substack{e\text{ long}\\ e\cap G_z\ne\varnothing}}(1-p_e)
\ge\exp\bigl[-44\beta sF_s-2\beta\eta^2\mathcal M(s,\lambda)\bigr].
\label{eq:smj-unbridged}
\end{equation}
For the first product, $1-\eta^2(1-e^{-t})\ge e^{-\eta^2t}$ by convexity of
the exponential, with $t=2\beta J(|e|)$, and the spanning coupling is at
most $\mathcal M(s,\lambda)$. For the second, the guard is touched by long
edges of total coupling at most $2\cdot11s\,F_s$. An edge appearing in both
products contributes two factors, which only lowers the bound; FKG is
applied to finitely many events and the infinite guard closure is reached
by a decreasing limit. Only the spanning bonds carry the factor $\eta^2$;
the guard cost $44\beta sF_s$ does not.

\subsection{Exposure and the recursion at every real scale}
\label{sec:smj-recursion}

\textit{A crossing at scale $\lambda s$ must cross every unbridged
candidate.} Let a $\lambda s$-crossing of $I_{\lambda s}$ be given, and
erase its loops, so that it is a self-avoiding open path
$v_0,\dots,v_k$ with $v_0<-3\lambda s$, $v_k\ge3\lambda s$, intermediate
sites in $I_{\lambda s}$, and edges of length at most $\lambda s$. Let
$z+I_s$ be an unbridged candidate. Let $i$ be the first index with
$v_i\ge z+3s$ and $j<i$ the last index with $v_j<z-3s$; the sites strictly
between lie in the block. If $i>j+1$, every edge of $v_j,\dots,v_i$ touches
the block, hence the guard, and is short by (a): this subpath is an
$s$-crossing. If $i=j+1$, the edge $\{v_j,v_i\}$ spans the block and has
length at most $\lambda s$. The two pieces $v_j,\dots,v_0$ and
$v_i,\dots,v_k$ do not use it and end outside $I_{\lambda s}$, at distance
more than $R$ from $v_j$ and $v_i$, so each leaves the radius-$R$ box of
its endpoint: the edge is a bridge, which (b) forbids. Only the order of
sites on the line and the allowed lengths are used, not where $J$ is
positive.

\textit{Exposure.} Reveal first every long edge touching $I_{\lambda s}$,
and then every short edge touching the box $v+\Lambda_R$ of an endpoint
$v$ of a revealed open long edge. Call the revealed information
$\mathcal F$. Whether a candidate is unbridged is $\mathcal F$-measurable:
(a) involves long edges only, and a bridge is an open long edge whose
endpoint boxes have been revealed. No short edge touching an unbridged
candidate is revealed. Such an edge would have an endpoint within $R$ of an
endpoint $v$ of an open long edge and within $s$ of the block, putting $v$
within $s+R<2s$ of the block, hence in $G_z$, which (a) forbids.

\textit{Sequential domination.} Let $\mathcal U$ be the set of unbridged
candidates, ordered from left to right. Conditional on $\mathcal F$ and on
the short edges touching the earlier members of $\mathcal U$, the short
edges touching the next member $B$ are still unrevealed, and the long edges
touching $B$ are known closed. Conditioning further on all other edges not
touching $B$, the edges touching $B$ follow a boundary law at $B$ with the
long ones absent, so $B$ is crossed with probability at most $1-u(s)$.
Multiplying,
\begin{equation}
\phi\bigl(\text{every member of }\mathcal U\text{ is }s\text{-crossed}
\,\big|\,\mathcal F\bigr)\le\bigl(1-u(s)\bigr)^{|\mathcal U|}.
\label{eq:smj-product}
\end{equation}

\textit{The recursion.} By the deterministic implication and
Eq.~\eqref{eq:smj-product}, after $\bar\Lambda\uparrow\ZZ$,
$u(\lambda s)\ge1-\EE\bigl[(1-u(s))^{|\mathcal U|}\bigr]$. Put
$t=\min\{u(s),\lambda^{-1}\}$. Since $|\mathcal U|\le\kappa\le\lambda$, we have
$t|\mathcal U|\le1$, and $1-e^{-x}\ge x/e$ on $[0,1]$, so
$1-(1-u(s))^{|\mathcal U|}\ge1-e^{-t|\mathcal U|}\ge e^{-1}t|\mathcal U|$.
Each of the $\kappa\ge\lambda/10$ candidates is unbridged with at least the
probability~\eqref{eq:smj-unbridged}, which bounds $\EE|\mathcal U|$ from below; taking
expectations,
\begin{equation}
u(\lambda s)\ge\frac{\lambda}{10e}\,
\exp\bigl[-44\beta sF_s-2\beta\eta^2\mathcal M(s,\lambda)\bigr]\,
\min\{u(s),\lambda^{-1}\}
\qquad(s\ge s_0,\ \lambda\ge36).
\label{eq:smj-recursion}
\end{equation}
This holds at every real $s\ge s_0$, not only along a geometric sequence,
and involves no interpolation between integer scales. For the
inverse-square kernel $sF_s$ and $\mathcal M(s,\lambda)/\ln\lambda$ are
bounded, and Eq.~\eqref{eq:smj-recursion} iterated along $s_0\lambda^n$
reproduces the inverse-square argument of Ref.~\cite{Duminil-Copin_2024lr}; here they are
not bounded, and we average instead.

\subsection{Termination estimate}
\label{sec:smj-termination}

Let $A_s$ be the event that a site of $I_s$ is connected to a site outside
$[-10s,10s)$. Let $D_s$ be the event that every long edge touching
$[-10s,10s)\cap\ZZ$ is closed; that interval has at most $21s$ sites,
so by (ii) $\phi^w(D_s)\ge e^{-84\beta sF_s}$. Consider the two side blocks
$\pm z_0+I_s$ with $z_0=\lceil6s\rceil$. They lie outside $I_s$, since
$z_0-3s\ge3s$, and inside $[-10s,10s)$, since $z_0+3s\le9s+1\le10s$. If
$D_s$ occurs and neither side block is $s$-crossed, then $A_s$ fails: a
path from $I_s$ to the outside, followed to its first exit from
$[-10s,10s)$, uses only short edges, starts on the inner side of one side
block and ends beyond its outer side, so by the argument of
Sec.~\ref{sec:smj-recursion} it contains an $s$-crossing of that block (a
single edge spanning the block would be long). The three events are
decreasing, each side block is uncrossed with probability at least $u(s)$
under $\phi^w$, and FKG gives
\begin{equation}
1-\phi^w(A_s)\ge e^{-84\beta sF_s}\,u(s)^2 .
\label{eq:smj-termination}
\end{equation}

Now use order. By Eq.~\eqref{eq:smj-zeroone} an infinite cluster exists
almost surely; the events that $I_s$ meets an infinite cluster increase to
$\mathcal I$ as $s\to\infty$, and each is contained in $A_s$. Hence
$\phi^w(A_s)\to1$, and $L(s)=-\ln[1-\phi^w(A_s)]$, finite by
Eq.~\eqref{eq:smj-termination}, tends to infinity. In terms of $q$,
\begin{equation}
2q(s)+84\beta sF_s\ge L(s)\xrightarrow[s\to\infty]{}\infty .
\label{eq:smj-L}
\end{equation}
The block crossings must become rare, or the tail budget must become large;
this is what converts the recursion into a contradiction.

\subsection{Logarithmic averages}
\label{sec:smj-average}

Write $E(x)=\sum_r\min(r,x)J(r)$ for real $x$. Since $E$ is nondecreasing,
$E(\lfloor x\rfloor)\le E(x)\le E(\lceil x\rceil)$, and
$\ln\lceil x\rceil/\ln x\to1$, so
$\limsup_{x\to\infty}E(x)/\ln x=C_+$. The two costs of
Eq.~\eqref{eq:smj-recursion} average over $\ln s$ to the domain energy.

\textit{Tail budget.} Exchanging sum and integral (all terms are
nonnegative), for $X\ge s_0$,
\begin{equation}
\int_{s_0}^{X}sF_s\,\frac{ds}{s}
=\sum_rJ(r)\int_{s_0}^{X}\one_{\{s<r\}}\,ds
=E(X)-E(s_0).
\label{eq:smj-intB}
\end{equation}
This is an identity: every pulse of the coupling, including those at
distances beyond $X$, is charged exactly once.

\textit{Spanning weight.} The atom $rJ(r)$ enters $\mathcal M(s,\lambda)$
only for $r/\lambda\le s<r$, a range of logarithmic length $\ln\lambda$,
and not at all if $r>\lambda X$. Hence
\begin{equation}
\int_{s_0}^{X}\mathcal M(s,\lambda)\,\frac{ds}{s}
\le\ln\lambda\sum_{r\le\lambda X}rJ(r)\le(\ln\lambda)\,E(\lambda X).
\label{eq:smj-intM}
\end{equation}
Divided by $\ln X$, the right sides of Eqs.~\eqref{eq:smj-intB}
and~\eqref{eq:smj-intM} have upper limits $C_+$ and $C_+\ln\lambda$.

\textit{The cap correction averages to zero.} Fix $\lambda$, put
$a=\ln\lambda$ and $g(s)=\bigl(a-q(s)\bigr)_+$. Then
\begin{equation}
\lim_{X\to\infty}\frac1{\ln X}\int_{s_0}^{X}g(s)\,\frac{ds}{s}=0 .
\label{eq:smj-cap}
\end{equation}
Fix $M_0>0$. By Eq.~\eqref{eq:smj-L} there is $s_1$ with
$L(s)>2a+84\beta M_0$ for $s\ge s_1$. For such $s$, $q(s)<a$ forces
$84\beta sF_s>84\beta M_0$, so
$g(s)\le a\one_{\{sF_s>M_0\}}\le(a/M_0)\,sF_s$. On $[s_0,s_1]$,
$g\le a$ contributes a constant. By Eq.~\eqref{eq:smj-intB} the upper limit
in Eq.~\eqref{eq:smj-cap} is at most $aC_+/M_0$, and $M_0$ is arbitrary.
No bound on $sF_s$ at individual scales is used.

\subsection{The final inequality and the gap}
\label{sec:smj-final}

Take $-\ln$ of Eq.~\eqref{eq:smj-recursion}. Since
$-\ln\min\{u,\lambda^{-1}\}=\max\{q,a\}=q+g$,
\begin{equation}
q(\lambda s)-q(s)\le-\ln\lambda+\ln(10e)+44\beta sF_s
+2\beta\eta^2\mathcal M(s,\lambda)+g(s)\qquad(s\ge s_0).
\label{eq:smj-drift}
\end{equation}
Integrate against $ds/s$ over $[s_0,X]$. Substituting $t=\lambda s$ in the
first term, the left side becomes
\begin{equation}
\int_{X}^{\lambda X}q(t)\,\frac{dt}{t}-\int_{s_0}^{\lambda s_0}q(t)\,\frac{dt}{t}
\ \ge\ -\int_{s_0}^{\lambda s_0}q(t)\,\frac{dt}{t},
\label{eq:smj-telescope}
\end{equation}
because $q\ge0$; the remaining constant is finite by
Eq.~\eqref{eq:smj-qbound}. The growing upper end needs no bound. Dividing
by $\ln X$, letting $X\to\infty$ and using
Eqs.~\eqref{eq:smj-intB}--\eqref{eq:smj-cap},
\begin{equation}
0\le-\bigl(1-2\beta\eta^2C_+\bigr)\ln\lambda+\ln(10e)+44\beta C_+ .
\label{eq:smj-final}
\end{equation}

Suppose now that $m>0$ and $2\beta C_+m^2<1$. Because $m<1$ we may choose
$\eta\in(m,1)$ with $2\beta C_+\eta^2<1$ (any $\eta$ if $C_+=0$). With $R$
and $s_0$ fixed by $\eta$ as in Sec.~\ref{sec:smj-escape}, choose
$\lambda\ge36$ so large that
$(1-2\beta\eta^2C_+)\ln\lambda>\ln(10e)+44\beta C_+$. This contradicts
Eq.~\eqref{eq:smj-final}. The order of choices is $\beta$, $m$, $\eta$, $R$,
$s_0$, $\lambda$, then $X\to\infty$; Eq.~\eqref{eq:smj-cap} is established
at fixed $\lambda$, with its own limit $M_0\to\infty$, before it is used. At
$\beta=0$ every edge is closed and $m_+=0$. This proves
Eq.~\eqref{eq:smj-gap}, and with Eq.~\eqref{eq:smj-mless1},
Eq.~\eqref{eq:smj-nobelow}.

Only the cumulative domain energy enters. The rare large values of $sF_s$
are paid for once through the identity~\eqref{eq:smj-intB}; the fixed guard
cost is divided by $\ln\lambda$ in Eq.~\eqref{eq:smj-final}, not
multiplied by $\eta^2$; and the spanning weight carries the factor
$\eta^2$ that makes small magnetization self-defeating.

\section{The jump at the critical point}
\label{sec:smj-endpoint}

This section proves part~(ii) of Theorem~2 of the main text. Let $C_+<\infty$ and suppose that $m_+(\beta_0)>0$ at one finite
$\beta_0$. For $0<C_-\le C_+<\infty$ such a point is supplied by
Theorem~1 of the main text. Then
\begin{equation}
0<\beta_c<\infty,\qquad
m_+(\beta_c)\ge\bigl(2\beta_cC_+\bigr)^{-1/2}>0,\qquad
\beta_c>\frac1{2C_+}.
\label{eq:smj-jump}
\end{equation}

\textit{The critical point is positive and finite.} Order at $\beta_0$
forces $\Sigma_J>0$, and by Eq.~\eqref{eq:smj-gap} also $C_+>0$. The Gibbs
measure is unique, and hence $m_+=0$, whenever
$2\beta \Sigma_J<1$~\cite{Dobruschin_1968td}, so
$1/(2\Sigma_J)\le\beta_c\le\beta_0$.

\textit{The bound holds above $\beta_c$.} The magnetization $m_+(\beta)$ is
nondecreasing in $\beta$~\cite{Griffiths_1967ci,Kelly_1968gg}, so every
$\beta>\beta_c$ is ordered: if some $\beta>\beta_c$ had $m_+(\beta)=0$, all
smaller $\beta$ would too, and the infimum would be at least $\beta$. For a
sequence $\beta_n\downarrow\beta_c$, Eq.~\eqref{eq:smj-gap} gives
$m_+(\beta_n)\ge(2\beta_nC_+)^{-1/2}$. The constant $C_+$ belongs to the
kernel; the auxiliary $\eta,R,s_0,\lambda$ of the proof may depend on $n$.

\textit{Right-continuity carries it to $\beta_c$.} For fixed $R$,
$\rho_R(\beta)=\langle\sigma_0\rangle^+_{\Lambda_R,\beta}$ is a finite sum of
exponentials of energies that include the exterior fields
$\sum_{j\notin\Lambda_R}J(|i-j|)\le2\Sigma_J$, so it is continuous in $\beta$.
Hence $m_+(\beta_n)\le\rho_R(\beta_n)\to\rho_R(\beta_c)$, and taking the
infimum over $R$,
$\lim_nm_+(\beta_n)\le m_+(\beta_c)$; this is the right-continuity of
Lemma~3.2 of Ref.~\cite{Aizenman_1988do}, with the limits taken in this
order. Therefore $m_+(\beta_c)\ge\lim_n(2\beta_nC_+)^{-1/2}
=(2\beta_cC_+)^{-1/2}$. Since $\beta_c$ is itself ordered,
Eq.~\eqref{eq:smj-nobelow} gives $\beta_c>1/(2C_+)$.

The infinite-volume magnetization is taken first and the temperature limit
second; no finite-volume quantity is evaluated at $\beta_c$. The proof of
Eq.~\eqref{eq:smj-jump} uses the ordering criterion only to produce
$\beta_0$; the gap itself does not depend on it.

\section{Comparison with the pointwise inverse-square coefficient}
\label{sec:smj-pointwise}

Earlier jump theorems assume a finite pointwise coefficient
$\limsup_{r\to\infty}r^2J(r)$~\cite{Aizenman_1988do}. It controls the
cumulative one:
\begin{equation}
C_+\le\limsup_{r\to\infty}r^2J(r) .
\label{eq:smj-CTK}
\end{equation}
If this $\limsup$ is infinite there is nothing to prove. Otherwise, for every $L$ larger than it there is
$r_0$ with $J(r)\le L/r^2$ for $r>r_0$; then for $N>r_0$,
$S_N\le\sum_{r\le r_0}rJ(r)+L\sum_{r=2}^{N}r^{-1}\le O(1)+L\ln N$ and
$NF_N\le LN\sum_{r>N}r^{-2}\le L$, so $E_N\le L\ln N+O(1)$ and
$C_+\le L$. The gap~\eqref{eq:smj-gap} is therefore never weaker than its
pointwise counterpart, and it applies to couplings with $r^2J(r)$ unbounded, such as
the dyadic chain of Sec.~\ref{sec:smj-dyadic}.

\textit{Recovery of the inverse-square jump.} If $r^2J(r)\to c$, the same estimate with
$J(r)\ge(c-\epsilon)/r^2$ eventually gives $E_N/\ln N\to c$, so
$C_-=C_+=c$. If moreover $\sum_rr\,|J(r)-c/r^2|<\infty$, then
$E_N=c\ln N+O(1)$, since that sum bounds the change of $E_N$; for
$J(r)=1/r^2$ exactly, $E_N=\sum_{r\le N}r^{-1}+N\sum_{r>N}r^{-2}=\ln N+1+\gamma_E+o(1)$,
with $\gamma_E$ Euler's constant. Reference~\cite{Aizenman_1988do}
writes $\mathcal H=-\tfrac12\sum_{\{x,y\}}J'_{x-y}(\sigma_x\sigma_y-1)$
with each pair once [its Eq.~(1.1)], so
$J'=2J$ at the same $\beta$, and its bond parameter
$1-e^{-\beta J'(r)}$ equals $p_\beta(r)$. Its coefficient
$J^+=\lim r^2J'(r)$ equals $2c$. Its dichotomy ``$M=0$ or
$\beta J^+M^2\ge1$'' reads $m_+=0$ or $2\beta c\,m_+^2\ge1$, and its jump
$M(\beta_c)\ge(\beta_cJ^+)^{-1/2}$ reads
$m_+(\beta_c)\ge(2\beta_cc)^{-1/2}$. With $C_+=c$,
Eqs.~\eqref{eq:smj-gap} and~\eqref{eq:smj-jump} reproduce both exactly,
normalization included, and Eq.~\eqref{eq:smj-nobelow} reproduces the
absence of order for $2\beta c\le1$.

\section{The dyadic chain}
\label{sec:smj-dyadic}

The dyadic chain has one bond per octave of distance,
\begin{equation}
J_a(2^k)=a\ln2\;2^{-k}\quad(k\ge0),\qquad J_a(r)=0\ \text{otherwise},
\label{eq:smj-dyadic}
\end{equation}
with amplitude $a>0$, nearest-neighbor bond $J_a(1)=a\ln2$, and
$\Sigma_J=2a\ln2$.

\textit{Domain energy.} With $\ell=\lfloor\log_2N\rfloor$, the bonds
$2^k\le N$ contribute $2^kJ_a(2^k)=a\ln2$ each, and those beyond $N$
contribute $N\sum_{k>\ell}a\ln2\,2^{-k}=a\ln2\,N2^{-\ell}$:
\begin{equation}
E_N=a\ln2\Bigl(\ell+1+\frac{N}{2^\ell}\Bigr),\qquad
E_{2N}-E_N=a\ln2 .
\label{eq:smj-dyadicT}
\end{equation}
Every octave of length adds the same energy. Since
$\ell\ln2\in(\ln N-\ln2,\ln N]$ and $N/2^\ell\in[1,2)$,
$|E_N-a\ln N|\le3a\ln2$, so $C_-=C_+=a$. On the support,
$r^2J_a(r)=a\ln2\,2^k\to\infty$, so the pointwise
theorem does not apply.

\textit{Amplitude scaling.} Every coupling, the nearest-neighbor bond
included, is proportional to $a$, and so are the exterior fields of a
plus-boundary box. The finite-volume plus Gibbs weights therefore depend on
$a$ and $\beta$ only through $a\beta$, and after the volume limit
\begin{equation}
m_{+,a}(\beta)=m_{+,1}(a\beta),\qquad \beta_c(a)=\beta_c(1)/a .
\label{eq:smj-dyadicscale}
\end{equation}

\textit{Jump.} Since $C_-=a>0$, Theorem~1 of the main text gives
a finite ordered point, and Eq.~\eqref{eq:smj-jump} with $C_+=a$ gives
\begin{equation}
0<\beta_c(a)<\infty,\qquad
m_{+,a}\bigl(\beta_c(a)\bigr)\ge\bigl[2a\beta_c(a)\bigr]^{-1/2},\qquad
\beta_c(a)>\frac1{2a}.
\label{eq:smj-dyadicjump}
\end{equation}
By Eq.~\eqref{eq:smj-dyadicscale} these are one statement about
$a\beta_c(a)=\beta_c(1)$. The bound $\beta_c(a)>1/(2a)$ improves on the
uniqueness bound $\beta_c(a)\ge1/(2\Sigma_J)=1/(4a\ln2)\approx0.361/a$.

\section{The pulse coupling}
\label{sec:smj-pulse}

The pulse coupling is
\begin{equation}
J(1)=1,\qquad J(R_n)=\theta_n=\frac{2^n}{R_n},\quad R_n=2^{2^n}\ (n\ge1),
\qquad J(r)=0\ \text{otherwise}.
\label{eq:smj-pulse}
\end{equation}
The pulses sit at $4,16,256,65536,\dots$, with $R_{n+1}=R_n^2$ and
$\theta_{n+1}/\theta_n=2/R_n\le1/2$. Hence $\Sigma_J\le1+2\theta_1=2$, and the tails
$\Theta_m=\sum_{n\ge m}\theta_n$ satisfy $\theta_m\le\Theta_m\le2\theta_m$.

\begin{figure}[t]
\centering
\includegraphics[width=0.6\columnwidth]{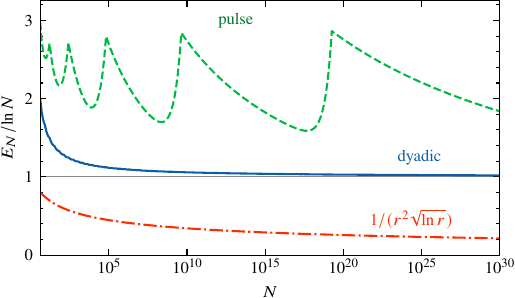}
\caption{Exact $E_N/\ln N$ for three couplings. The dyadic chain with $a=1$
(solid) tends to $C_-=C_+=1$: it orders, and its magnetization jumps. The
pulse coupling (dashed) oscillates between positive bounds: it also orders and
jumps. The log-corrected tail $J(r)=1/(r^2\sqrt{\ln r})$, $r\ge3$
(dash-dotted), tends to zero: no order at any finite temperature.}
\label{fig:sm-growth}
\end{figure}

\textit{Logarithmic domain energy.} For an integer $N\ge4$ choose $n\ge1$
with $R_n\le N<R_{n+1}$. The nearest-neighbor bond contributes $1$, each
saturated pulse $R_m\le N$ contributes $R_m\theta_m=2^m$, and the rest
contributes $NF_N=N\Theta_{n+1}$:
\begin{equation}
E_N=2^{n+1}-1+N\Theta_{n+1}.
\label{eq:smj-pulseT}
\end{equation}
For the lower bound, $\ln N<\ln R_{n+1}=2^{n+1}\ln2$ gives
$E_N\ge2^{n+1}-1\ge2^n>\ln N/(2\ln2)$. For the upper bound,
$2^{n+1}=2\ln R_n/\ln2\le2\ln N/\ln2$; and since $x/\ln x$ increases
for $x>e$ and $4\le N<R_{n+1}$, $(N/R_{n+1})\ln R_{n+1}\le\ln N$, so
$N\Theta_{n+1}\le2N\theta_{n+1}
=\frac{2}{\ln2}\frac{N}{R_{n+1}}\ln R_{n+1}\le\frac{2\ln N}{\ln2}$.
Hence for every integer $N\ge4$,
\begin{equation}
\frac{\ln N}{2\ln2}\le E_N\le\frac{4\ln N}{\ln2},
\qquad 0<\frac1{2\ln2}\le C_-\le C_+\le\frac4{\ln2}.
\label{eq:smj-pulsebounds}
\end{equation}
Figure~\ref{fig:sm-growth} compares the exact ratio $E_N/\ln N$ of the pulse
coupling with that of the dyadic chain of Sec.~\ref{sec:smj-dyadic} at $a=1$
and of the log-corrected tail $J(r)=1/(r^2\sqrt{\ln r})$, $r\ge3$.

\textit{Unbounded per-scale budgets.} At $s=R_n/2$ the pulse at $R_n>s$ lies
in the tail, so
\begin{equation}
sF_s\ge\frac{R_n}{2}\,\theta_n=2^{n-1}\to\infty .
\label{eq:smj-pulseB}
\end{equation}
The annular budget $\mathcal W(s,\lambda)=\sum_{6s<r\le\lambda s}(r-6s)J(r)$,
which up to the rounding of $6s$ is the weight of bonds of length at most
$\lambda s$ spanning a block of Sec.~\ref{sec:smj-blocks}, is unbounded at
fixed ratio as well. At $s=R_n/8$ and $\lambda=8$ ($n\ge2$), the annulus
$3R_n/4<r\le R_n$ contains the pulse $R_n$ and no other bond, because
$R_{n-1}=\sqrt{R_n}\le R_n/4$ and $R_{n+1}>R_n$, so
\begin{equation}
\mathcal W(R_n/8,8)=\Bigl(R_n-\frac34R_n\Bigr)\theta_n=2^{n-2}\to\infty .
\label{eq:smj-pulseW}
\end{equation}
For $n=2$ this is $\mathcal W(2,8)=4J(16)=1$. A logarithmic domain energy
therefore does not imply bounded costs at every scale: a single pulse can
use a logarithmic share of the cumulative budget at one scale. This is why
Sec.~\ref{sec:smj-gap} averages over $\ln s$ rather than iterating along
scales.

\textit{Order and jump.} Since $C_->0$, Theorem~1 of the main
text gives order at some finite temperature, and Eq.~\eqref{eq:smj-jump}
with $C_+\le4/\ln2$ gives
\begin{equation}
m_+(\beta_c)\ge\sqrt{\frac{\ln2}{8\beta_c}}>0 .
\label{eq:smj-pulsejump}
\end{equation}

\section{Support dilation: the domain energy does not fix the jump}
\label{sec:smj-dilation}

For an integer $d\ge2$ the dilation of $J$ is
\begin{equation}
J^{[d]}(r)=J(r/d)\ \text{if}\ d\mid r,\qquad J^{[d]}(r)=0\ \text{otherwise}.
\label{eq:smj-dilation}
\end{equation}
It is nonnegative and translation invariant, with the same total coupling
$\Sigma_J$.

\textit{Exact decoupling with the plus boundary kept.} Let
$I\subset\ZZ$ be a finite interval and
$H^+_{J,I}(\sigma)=-\sum_{i<j\in I}J(j-i)\sigma_i\sigma_j
-\sum_{i\in I}\sigma_i\sum_{j\notin I}J(|j-i|)$ its plus-boundary
Hamiltonian with all exterior couplings. For each residue
$\rho\in\{0,\dots,d-1\}$ let $I_\rho=\{k:\rho+dk\in I\}$, again an interval,
and $\tau^{(\rho)}_k=\sigma_{\rho+dk}$. The dilated coupling joins only sites
of the same residue, with $J^{[d]}(d|k-l|)=J(|k-l|)$, and the exterior field
at $i=\rho+dk$ is
$\sum_{j\notin I}J^{[d]}(|j-i|)=\sum_{l\notin I_\rho}J(|l-k|)$, since the
exterior sites coupled to $i$ are exactly $\rho+dl$ with $l\notin I_\rho$.
Hence
\begin{equation}
H^+_{J^{[d]},I}(\sigma)=\sum_{\rho=0}^{d-1}H^+_{J,I_\rho}(\tau^{(\rho)}),
\label{eq:smj-decouple}
\end{equation}
and the plus-boundary Gibbs measure of $J^{[d]}$ on $I$ is the product of
$d$ plus-boundary Gibbs measures of $J$ on the $I_\rho$. As $I$ exhausts
$\ZZ$, each $I_\rho$ does, so at every site
$i=\rho+dk$,
$\langle\sigma_i\rangle^+_{J^{[d]},I}=\langle\tau_k\rangle^+_{J,I_\rho}\to
m_{+,J}(\beta)$. The magnetization per physical site is unchanged, and so
is the critical point:
\begin{equation}
m_{+,J^{[d]}}(\beta)=m_{+,J}(\beta),\qquad
\beta_c(J^{[d]})=\beta_c(J).
\label{eq:smj-dilm}
\end{equation}

\textit{The domain energy is multiplied by $d$.} Relabeling the absolutely
convergent sum,
\begin{equation}
E_N(J^{[d]})=\sum_{k\ge1}\min(dk,N)J(k)=d\,E_{N/d}(J),\qquad
E_{dn}(J^{[d]})=d\,E_n(J),
\label{eq:smj-dilT}
\end{equation}
where $E_x$ at real $x$ means $E(x)$. Since
$E_{\lfloor N/d\rfloor}\le E(N/d)\le E_{\lceil N/d\rceil}$ and
$\ln(N/d)/\ln N\to1$, $C_-$ and $C_+$ are multiplied by $d$, and so is
the common value $c=C_-=C_+=\lim E_N/\ln N$ when it exists. The jump ratio
$Q_c=2\beta_c c\,m_+(\beta_c)^2$, which Eq.~\eqref{eq:smj-jump} bounds
below by one, therefore obeys
\begin{equation}
Q_c(J^{[d]})=d\,Q_c(J).
\label{eq:smj-dilQ}
\end{equation}

\textit{An explicit family with $Q_c\ge d$.} Take $J(r)=1/r^2$. It has
$c=1$ (Sec.~\ref{sec:smj-pointwise}), orders at low temperature by the
theorem of Fr\"ohlich and Spencer~\cite{Frohlich_1982tp}, and is disordered
for $2\beta \Sigma_J<1$, so $0<\beta_c<\infty$; Eq.~\eqref{eq:smj-jump} (or
Ref.~\cite{Aizenman_1988do}) gives $Q_c\ge1$. Its dilation
\begin{equation}
J^{[d]}(r)=\frac{d^2}{r^2}\ \text{if}\ d\mid r,\qquad 0\ \text{otherwise},
\label{eq:smj-dilexample}
\end{equation}
has $E_N=d\ln N+O(1)$, the same $\beta_c$ and the same $m_+(\beta_c)$, and
\begin{equation}
Q_c(J^{[d]})=d\,Q_c(1/r^2)\ge d>1 .
\label{eq:smj-dilQd}
\end{equation}
No universal equality $Q_c=1$ holds on the class $0<C_-=C_+<\infty$, and
$Q_c$ has no uniform upper bound there. The mechanism is visible in
Eq.~\eqref{eq:smj-decouple}: the domain energy counts every broken bond, but
after dilation each sublattice feels only its own bonds, so $C_+$ decides
whether a jump must occur, not its size. Correspondingly, the gap applied
to $J^{[d]}$ directly gives $m_+(\beta_c)\ge(2\beta_cd)^{-1/2}$, weaker by
$\sqrt d$ than the bound $(2\beta_c)^{-1/2}$ inherited from
Eq.~\eqref{eq:smj-dilm}. Whether $Q_c=1$ for kernels whose support
generates $\ZZ$, for the dyadic chain, or for $1/r^2$ itself is not
decided.

\end{document}